\documentclass[a4paper,11pt]{article}

\usepackage{jheppub} 

\usepackage[T1]{fontenc} 

\title{\boldmath Constraining bulk-to-boundary correlators  under Poincar\'e symmetry}

\author{Jiang Long and}
\author{Jing-Long Yang}

\affiliation{School of Physics, Huazhong University of Science and Technology, \\ Luoyu Road 1037, Wuhan, Hubei 430074, China}

\emailAdd{longjiang@hust.edu.cn}
\emailAdd{yangjinglong@hust.edu.cn}

\abstract{It is well known that a general two-point function cannot be uniquely determined by Poincar\'e symmetry. In this paper, we show that bulk-to-boundary correlators are highly constrained after imposing suitable fall-off conditions near future/past null infinity. More precisely, scalar bulk-to-boundary correlators are fixed to a unique form up to a normalization constant, whereas fermionic bulk-to-boundary correlators are fixed to a linear superposition of scalar and fermionic branches. This is established by asymptotically expanding the  Ward identities, where upon the leading terms decouple from the subleading ones. In the fermionic branch, the power-law exponent of the bulk-to-boundary correlator is greater by one than the fall-off index. Consequently, we revisit the relation between Carrollian correlators and momentum space scattering amplitudes for fermionic operators. In this context, we find that the Fourier transform bridging the two acquires an extra factor of $\sqrt{\omega}$ for each fermionic operator.  Furthermore, we reduce the bulk-to-boundary correlator to the boundary-to-boundary correlator and identify a critical fall-off index $\Delta=1$. For $0 < \Delta < 1$, only a magnetic branch exists for scalars. For $\Delta > 1$, the electric branch is always divergent for both scalar and fermionic branches and thus requires regularization.}

\begin{document} 
\maketitle
\flushbottom

\section{Introduction}
Poincar\'e symmetry is fundamental to relativistic quantum field theory (QFT) \cite{1995qtf..book.....W}. However, it is insufficient to uniquely fix two-point correlation functions, a fact that leads to rich physical phenomena. This limitation motivates its extension to conformal symmetry, from which one obtains the broader class of conformal field theories (CFTs). Within CFTs, global conformal symmetry strongly constrains the structures of two- and three-point correlators \cite{Polyakov:1970xd,1974Non,Ferrara:1973eg}. 

Recently, the CFT technology has been adapted to the theories with Carrollian conformal symmetry and merged with the program of flat holography\cite{Polchinski:1999ry,Susskind:1998vk,Giddings:1999qu,Balasubramanian:1999ri,deBoer:2003vf,Gary:2009ae}. It is understood that the Poincar\'e symmetry and even the  BMS symmetry \cite{Bondi:1962px,Sachs:1962wk,PhysRev.128.2851,Barnich:2010eb} is a typical Carrollian conformal  symmetry \cite{Duval:2014uva,Duval:2014lpa,Duval:2014uoa,Bergshoeff:2017btm,Hartong:2015xda} at the null boundary of an asymptotically flat spacetime. The reason why one can use the CFT method in a theory with Poincar\'e symmetry lies in two aspects.  At first, the Poincar\'e group is the ultra-relativistic limit (where the velocity of light tends to 0) of the three-dimensional relativistic conformal group $SO(2,3)$ \cite{levy:1965} . Second, the Lorentz group, a subgroup of the Poincar\'e group, is  isomorphic to the two-dimensional global conformal group. The former leads to the Carrollian holography \cite{Bagchi:2025vri,Nguyen:2025zhg,Ruzziconi:2026bix} and the latter lies at the heart  of celestial holography \cite{Pasterski:2021raf}. Various results have been obtained by using Poincar\'e symmetry to constrain the boundary correlators. These include \cite{Bagchi:2016bcd,Chen:2021xkw,Nguyen:2023miw,Salzer:2023jqv} for Carrollian correlator, and \cite{Law:2019glh} for celestial correlator.

The fact that two-point bulk correlators are largely unconstrained by Poincar\'e symmetry, while their boundary counterparts are highly constrained by the same symmetry, presents a conceptual puzzle. This disparity highlights a gap in our understanding of bulk-to-boundary correlators, which serve as the crucial missing link between bulk-to-bulk and boundary-to-boundary ones. Indeed, bulk-to-boundary propagators\footnote{In this work, we use the term ``bulk-to-boundary propagator'' specifically for elementary operators. The more general term ``bulk-to-boundary correlator'' applies to all operators, including the composite operators. For instance, $\Phi$ is an elementary field in $\Phi^3$ theory, whereas $\Phi^2$ is composite. Our results are valid for both types.} are the fundamental building blocks for holographic correlators in both AdS/CFT \cite{Maldacena:1997re,Gubser:1998bc,1998AdTMP...2..253W,Banks:1998dd}  and Carrollian holography. To be more precise, the holographic Carrollian correlator, which is also called Carrollian amplitude \cite{Donnay:2022aba,Donnay:2022wvx,Bagchi:2022emh,Mason:2023mti} can be computed perturbatively using the bulk-to-boundary propagator, bulk-to-bulk propagator as well as interaction vertices \cite{Liu:2024nfc}. The same result can be obtained by taking the flat limit of AdS/CFT correlators \cite{Alday:2024yyj}. The results have been extended to higher dimensions \cite{Liu:2024llk,Kulkarni:2025qcx}, Rindler \cite{Li:2024kbo}  and thermal spaces \cite{Long:2025bfi}. Further recent developments can be found in \cite{Chakrabortty:2024bvm,Adamo:2025bfr,Lipstein:2025jfj,Nguyen:2025sqk}. 

In this work, we investigate and constrain the form of bulk-to-boundary correlators in theories with Poincar\'e symmetry. Remarkably, after imposing appropriate fall-off conditions near future/past null infinity, these correlators are largely fixed, admitting only a limited number of possible structures.

 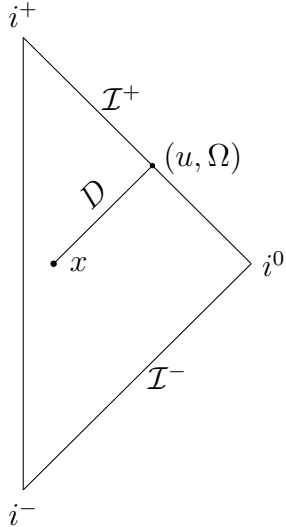
\begin{figure}
    \begin{center}
\begin{tikzpicture}[scale=1.0,>=stealth]

\coordinate (iplus) at (0,3);    
\coordinate (i0)   at (3,0);    
\coordinate (iminus) at (0,-3); 

\draw (iplus) -- (i0) node[midway,right,xshift=-0.6cm,yshift=0.7cm] {$\mathcal{I}^+$};
\draw (i0) -- (iminus) node[midway,right] {$\mathcal{I}^-$};
\draw (iminus) -- (iplus);

\node[above] at (iplus) {$i^+$};
\node[right] at (i0)   {$i^0$};
\node[below] at (iminus) {$i^-$};

\filldraw (0.4,0) circle (1pt) node[anchor=west, xshift=2pt] {$x$};

\coordinate (sigmaJ) at (1.7,1.3); 
\filldraw (sigmaJ) circle (0.8pt) node[right,yshift=0.1cm] {$(u,\Omega)$}; 

\draw (0.4,0) -- (sigmaJ) node[midway,above,sloped,xshift=0.1cm] {$D$};

\end{tikzpicture}
\caption{The bulk-to-boundary correlator $D$. The pole of the bulk-to-boundary correlator should be located on the light ray  that connects a bulk field located at $x$ and a boundary operator located at $(u,\Omega)\in\mathcal I^+$. In general, this is a function of the translation invariant variable $\widehat u=u+n\cdot x$. }\label{fig1}
\end{center}\end{figure}

 More precisely, the bulk-to-boundary correlator is determined by the fall-off index $\Delta$ of the bulk fields and we identify two branches for the bulk-to-boundary correlators. As illustrated in figure \ref{fig1}, for a bulk field located at $x$ and a boundary operator located at $(u,\Omega)\in\mathcal I^+$, the connecting light ray allows the construction of a translation invariant quantity $\widehat u=u+n\cdot x$.  In the scalar branch, the bulk-to-boundary correlator is a power-law function of $\widehat u$ with an exponent equal to the fall-off index. In the fermionic branch, it remains a power-law function of $\widehat u$, but its exponent exceeds the fall-off index by one.
This property of the fermionic branch modifies the relationship between the Carrollian correlator and the momentum space scattering amplitude. As  suggested by \cite{Donnay:2022wvx}, the Carrollian correlator is a Fourier transform of the scattering amplitude $\mathcal A$
\bea &&\langle \prod_{j=1}^{m}\Sigma_{s_j}(u_j,\Omega_j,\sigma_j)\rangle=\prod_{j=1}^{m} \int d\omega_j e^{-i\sigma_j\omega_j u_j}\mathcal{A}_{s_1s_2\cdots s_j}( p_1,p_2,\cdots, p_{m}).
\eea However, for fermionic operators, this relation is corrected to 
\bea 
\langle \prod_{j=1}^m \uppsi_{s_j}(u_j,\Omega_j,\sigma_j,a_j)\rangle= \left(\prod_{j=1}^m \int_0^\infty d\omega_j \sqrt{\omega_j} e^{-i\sigma_j\omega_j u_j}\right)\mathcal A_{s_1s_2\cdots s_m}( p_1,p_2,\cdots,p_{m}).
\eea 
 It follows that the holographic Carrollian correlators involving fermions are automatically finite. Furthermore, there will be no magnetic type two-point holographic Carrollian correlator in the fermionic branch.  

The layout of this work is as follows. In section \ref{sectionw}, we introduce the notations and conventions in this paper and solve the Ward identities for the scalar and fermionic bulk-to-boundary correlators.The distinct nature of the fermionic solution leads us to revisit, in section \ref{cmpl}, the precise relationship between fermionic Carrollian amplitudes and momentum space scattering amplitudes. Subsequently, in section \ref{el}, we perform the reduction from bulk-to-boundary to boundary-to-boundary correlators, analyzing their detailed properties separately in the scalar and fermion branches. We conclude with a summary and a discussion of future directions in the final section.

\section{Ward identities for the bulk-to-boundary correlators}\label{sectionw}
In this section, we  derive the Ward identities for the bulk-to-boundary correlators in a Poincar\'e invariant theory. 
First, we introduce the conventions and notation used throughout. We  work in four-dimensional Minkowski spacetime $\mathbb R^{1,3}$ with the signature $(-,+,+,+)$. The Cartesian coordinates are denoted as $x^\mu=(t,x^i)$ and they are transformed to the retarded coordinates by the relation 
\bea 
x^\mu=u \bar m^\mu+r n^\mu
\eea where $r$ is the spatial radius in spherical coordinates $(r,\theta,\phi)$ and $u=t-r$ is the retarded time. The  $\bar m^\mu=(1,0,0,0)$ is a timelike unit vector while $n^\mu=(1,n^i)$ is a null vector with $n^i=(\sin\theta\cos\phi,\sin\theta\sin\phi,\cos\theta)$. There is another null vector $\bar n^\mu=(-1,n^i)$ whose inner product with $n^\mu$ is $2$. A pure spatial vector $m^\mu=(0,n^i)$ is also useful in the following. Obviously, $m^\mu$ and $\bar m^\mu$ are determined by the null vectors $n^\mu$ and $\bar n^\mu$
\be 
m^\mu=\frac{1}{2}(n^\mu+\bar n^\mu),\quad \bar m^\mu=\frac{1}{2}(n^\mu-\bar n^\mu).
\ee The spacetime translation operator $\partial_\mu$ can be expressed as 
\be 
\partial_\mu =-n_\mu \partial_u+m_\mu \partial_r-\frac{1}{r}Y_\mu^A\partial_A,
\ee where $Y_\mu^A=-\partial^A n_\mu$. The Capital Latin index $A=\theta,\phi$ is raised by the inverse metric $\gamma^{AB}$ of $S^2$ with 
\be 
\gamma_{AB}=\left(\begin{array}{cc}1&0\\ 0&\sin^2\theta\end{array}\right).
\ee The four-dimensinoal version of $\gamma_{AB}$ is 
\be 
\gamma_{\mu\nu}=\gamma_{AB}Y^A_\mu Y^B_\nu
\ee which is also related to the Minkowski metric through the identity 
\be 
\gamma_{\mu\nu}=\eta_{\mu\nu}-\frac{1}{2}(n_\mu \bar n_\nu+n_\nu \bar n_\mu).
\ee 
We can also define the advanced coordinates $(v,r,\theta,\phi)$ via
\be 
x^\mu=v\bar m^\mu+r\bar n^\mu
\ee and thus the spacetime translation operator becomes
\be 
\partial_\mu=\bar n_\mu \partial_v+m_\mu \partial_r-\frac{1}{r}Y_\mu^A\partial_A.
\ee 

\subsection{Scalar}
A scalar operator $\Phi(x)$ in a Poincar\'e invariant theory satisfies the transformation law
\be 
\Phi'(x')=\Phi(x),
\ee where $x'^\mu$ is 
\be 
x'^\mu=\Lambda^\mu_{\ \nu}x^{\nu}+c^\mu.
\ee The constant vector $c^\mu$ denotes the spacetime translation while the $4\times 4$ matrix $\Lambda^\mu_{\ \nu}$ represents the Lorentz transformation. We focus on the two-point correlator in the bulk at first:
\be 
G(x;y)=\langle \text{T}\Phi(x)\Phi(y)\rangle
\ee where $\text{T}$ is the time-ordering symbol
\be 
\text{T}\Phi(x)\Phi(y)=\theta(x^0-y^0)\Phi(x)\Phi(y)+\theta(y^0-x^0)\Phi(y)\Phi(x).
\ee The Ward identities for the correlator $G(x;y)$ are 
\bs\label{Wardbulk}\begin{align}
    \left(\partial_\mu^x +\partial_\mu^y\right)G(x;y)&=0,\\
    \left(x_{[\nu}\partial_{\mu]}^x+y_{[\nu}\partial_{\mu]}^y\right)G(x;y)&=0.
\end{align}\es The symbol $\partial_\mu^x$ denotes the partial derivative with respect to $x^\mu$:
\be
\partial_\mu^x=\frac{\partial}{\partial x^\mu}.
\ee
The same notation applies to the coordinate $y^\mu$. Square brackets $[\cdots]$ denote total antisymmetrization with unit normalization over the enclosed indices.
The first line of \eqref{Wardbulk} corresponds to the spacetime translation invariance of the theory while the second line stems from the Lorentz invariance. The solution of these ten Ward identities is 
\be 
G(x;y)=g(h)\label{bulktobulks}
\ee where \be h=-(x-y)^2\label{defh}\ee  is the Lorentz invariant spacetime interval between the two points. This appears to exhaust the constraints on the two-point correlator. 
To solve the Ward identities, we must also impose the fall-off conditions. In this work, we assume the existence of a boundary Carrollian field theory, where the scalar operator $\Phi(x)$ admits an asymptotic expansion
\bea 
\Phi(x)=\left\{\begin{array}{cc} \frac{\Sigma(u,\Omega)}{r^{\Delta}}+\cdots,& \text{near}\  \mathcal I^+,\\
\frac{\Sigma^{(-)}(v,\Omega)}{r^{\Delta}}+\cdots,&\text{near}\  \mathcal{I}^-.\end{array}\right.\label{falloffp}
\eea 
The operators $\Sigma(u,\Omega)$ and $\Sigma^{(-)}(v,\Omega)$ are understood as the fundamental fields at the future/past null infinity, respectively. We denote the angular coordinates $(\theta,\phi)$ collectively by $\Omega$. The constant $\Delta$, which characterizes the leading fall-off behavior of the bulk field, is called the fall-off index. To avoid divergence in large distance, the fall-off index should be non-negative. The value $\Delta=0$ is also allowed. This occurs, for instance, in theories exhibiting spontaneous symmetry breaking, where a nonzero vacuum expectation value for the order parameter leads to $\Delta=0$. Our work excludes such cases and focuses on the regime $\Delta>0$.
Two kinds of bulk-to-boundary correlators are defined as follows
\bs\begin{align}
    D(u,\Omega;x')&=\langle \Sigma(u,\Omega)\Phi(x')\rangle,\\
    D^{(-)}(v,\Omega;x')&=\langle \Phi(x')\Sigma^{(-)}(v,\Omega)\rangle.
\end{align}\es They are related to the bulk-to-bulk correlator via the bulk-boundary dictionary 
\bs\label{asymp}\begin{align}
    D(u,\Omega;x')&=\lim{}_+ r^\Delta G(x;x'),\\
    D^{(-)}(v',\Omega';x)&=\lim{}_{-}r^{\Delta}G(x;x').
\end{align} 
\es The limits  are defined as 
\bea 
\lim{}_+(\cdots)=\lim_{r\to\infty,\ u\ \text{finite}}(\cdots),\qquad \lim{}_-(\cdots)=\lim_{r\to\infty,\ v\ \text{finite}}(\cdots).
\eea 
We will mainly discuss the bulk-to-boundary correlator $D(u,\Omega;x')$. The bulk-to-bulk correlator may be expanded asymptotically by noticing the relation \eqref{asymp}
\be 
G(x;x')=\frac{D(u,\Omega;x')}{r^\Delta}+\text{subleading terms}.
\ee Therefore, we can find an asymptotic expansion for the Ward identities \eqref{Wardbulk}. At the leading order, they are 
\bs\label{Wardbtb}\begin{align}
    \partial_\mu' D(u,\Omega;x')&=n_\mu \partial_u D(u,\Omega;x'),\label{transW}\\
    x'_{[\nu}\partial_{\mu]}' D(u,\Omega;x')&=\left(-\frac{1}{2}n_{\mu\nu}u\partial_u-\frac{\Delta}{2}n_{\mu\nu}+\frac{1}{2}Y_{\mu\nu}^A \partial_A\right)D(u,\Omega;x').\label{LorentzW}
\end{align}\es 
We have defined  antisymmetric tensors
\be 
n^{\mu\nu}=n^{[\mu}\bar n^{\nu]},\quad Y_{\mu\nu}^A=Y_\mu^A n_\nu-Y_\nu^A n_\mu.
\ee 
Note that the leading order Ward identities only depend on the bulk-to-boundary correlator. In other words, the Ward identities for the bulk-to-boundary correlator decouple from the subleading ones. The first equation \eqref{transW} governs the spacetime translation invariance of the bulk-to-boundary correlator
whose solution is 
\be 
D(u,\Omega;x')=F(u+n\cdot x';\Omega).
\ee Substituting into the second equation \eqref{LorentzW}, we find \bea 
-\frac{1}{2}n_{\mu\nu}(u+n\cdot x')F'-n_{\mu\nu}\frac{\Delta}{2}F+n_{[\nu}Y_{\mu]}^A \delta_A F=0\label{deltaF}
\eea where 
\be 
F'=\frac{d}{du}F(u+n\cdot x';\Omega),\quad \delta_A F=\frac{\partial}{\partial\theta^A}F(u+n\cdot x';\Omega)
\ee with $\theta^A=(\theta,\phi)$. Note that \be Y^\mu_A\cdot n_\mu=Y^\mu_A \bar n_\mu=0,\ Y^\mu_A Y_{\mu B}=\gamma_{AB},\ee we multiply both sides of \eqref{deltaF} by $Y^\mu_B$ and find 
\be 
\delta_AF=0.
\ee Therefore, $F$ only depends on the variable $\widehat u=u+n\cdot x'$. The equation \eqref{deltaF} can be solved 
\bea 
D(u,\Omega;x')=F(u+n\cdot x')=\frac{C_s}{(u+n\cdot x')^\Delta}\label{bulktoboundary}
\eea where $C_s$ is a normalization constant whose subscript is a shorthand of scalar. Therefore, the two-point bulk-to-boundary correlator is completely determined by the Poincar\'e invariance. Our result is consistent with \cite{Chen:2023naw}. This is a bit confusing since the bulk-to-bulk correlator $G(x;x')$ has a large number of degrees of freedom, see \eqref{bulktobulks}. To understand this result, we notice that the fall-off condition \eqref{falloffp} impose strong constraints on the theory. At first, the condition \eqref{falloffp} excludes the massive theory since a mass term would violates the power-law  fall-off conditions. Physically, a massive particle can never reach $\mathcal I^+$ and thus the boundary field $\Sigma$ is ill-defined. To understand the massless feature of the bulk-to-boundary correlator, we may consider the poles  located 
at 
\be 
u+n\cdot x'=0.
\ee This is a null hypersurface that describes a massless particle moves from $x'$ to the boundary point $(u,\Omega)$.

Actually, the distance 
\bea 
h=-(x-x')^2=-(u\bar m+r n-x')^2=2r(u+n\cdot x')-(u\bar m-x')^2.
\eea Therefore, the bulk-to-bulk correlator $G(x;x')$ can only depend on two independent variables $r(u+n\cdot x')$ and $(u\bar m-x')^2$ in the asymptotic expansion.
At the leading order, the second term can be safely ignored and then 
\be 
G(x;x')=g(2r(u+n\cdot x'))+\text{subleading terms}.
\ee By assumption, the leading term of the function $g$ is fixed to the form $r^{-\Delta}$.  In the meanwhile, the variable $u+n\cdot x'$ always appears associated with $r$. Thus, 
\be 
G(x;x')\sim \frac{1}{r^\Delta (u+n\cdot x')^\Delta}+\text{subleading terms}.
\ee As a consequence, we find the same bulk-to-boundary correlator as \eqref{bulktoboundary}.

\textbf{Remarks.} Several points are listed in the following.
\begin{itemize}
    \item \textbf{Uniqueness.} Though the scalar bulk-to-boundary correlator is unique,  the two-point function in the bulk cannot  be determined by the bulk-to-boundary correlator in general.  As an illustration, the following two  bulk-to-bulk correlators 
    \bea 
    G^{(1)}(x;x')=\frac{1}{h^{\Delta}},\quad G^{(2)}(x;x')=\frac{1}{h^{\Delta}}+\frac{1}{h^{\Delta'}}
    \eea have the same leading fall-off  behavior near $\mathcal I^+$ for $\Delta'>\Delta$. Therefore, by extrapolating one of the points to the null boundary, the bulk-to-boundary correlator shares the same form.
    \item \textbf{The quantity $\widehat u$.} The variable $\widehat u$ appears in the bulk-to-boundary correlator since it is an invariant under spacetime translation \be 
    u\to u-c\cdot n,\quad \Omega\to \Omega,\quad x^\mu\to x^\mu+c^\mu,\quad \widehat u\to \widehat u.\label{translation}
    \ee Furthermore, this quantity transforms properly under Lorentz transformations, 
    \be 
    u\to \Gamma^{-1}u,\quad n^\mu\to \Gamma^{-1} {\Lambda^\mu}_\nu n^\nu,\quad x^\mu\to {\Lambda^\mu}_{\nu}x^\nu,\quad \widehat u\to \Gamma^{-1}\widehat u,\label{ntranslation}
    \ee making it suitable for bulk-to-boundary correlators. The factor $\Gamma$ is fixed by the Lorentz transformations whose form can be found in \cite{Liu:2024nfc}
    \bea 
\Gamma=\frac{|az+b|^2+|cz+d|^2}{1+z\bar z}=\frac{1+|z'|^2}{|a-cz'|^2+|b-dz'|^2}=\frac{(1+|z'|^2)|cz+d|^2}{1+|z|^2}
\eea  where we have adapted to the 
stereographic coordinates $(z,\bar z)$ which is related to the spherical coordinates by 
\be 
z=\cot\frac{\theta}{2}e^{i\phi},\quad \bar z=\cot\frac{\theta}{2}e^{-i\phi}.
\ee The constants $a,b,c,d$ characterize the M\"{o}bius transformation $SL(2,\mathbb{C})$ on $S^2$ \bea 
z'=\frac{az+b}{cz+d},\quad ad-bc=1.
\eea 

The bulk-to-boundary correlator is consistent with the Lorentz transformation law of the boundary field 
    \be 
    \Sigma(u,\Omega)\to \Sigma'(u',\Omega')=\Gamma^{\Delta}\Sigma(u,\Omega).
    \ee One can also construct various operators from $\Sigma$. Considering a linear operator $\mathscr L$ acting on $\Sigma$, then 
    \bea 
    \langle \mathscr L \Sigma(u,\Omega)\Phi(x)\rangle=\mathscr L D(u,\Omega;x').
    \eea In particular, 
    \bs\begin{align}
        \langle \dot\Sigma(u,\Omega)\Phi(x)\rangle&=-\frac{C_s\Delta}{(u+n\cdot x')^{\Delta+1}},\\ \langle \partial_A \Sigma(u,\Omega)\Phi(x)\rangle&=\frac{C_s\Delta}{(u+n\cdot x')^{\Delta+1}} Y_A\cdot x',
    \end{align}\es where $Y_A\cdot x'=Y_A^\mu x'_\mu$.
    \item \textbf{Fall-off conditions.} A more general fall-off condition involves a logarithm \footnote{In general relativity, the standard fall-off condition near $\mathcal I^{\pm}$ omits logarithmic terms ($\ln r$) and originates from the classic works of Bondi \cite{Bondi:1962px}, Sachs \cite{Sachs:1962wk}, and Penrose \cite{1965RSPSA.284..159P}. It is also natural, however, to consider spacetimes with metrics that allow expansions containing such terms \cite{Winicour1985LogarithmicAF}, as in \eqref{morefall}. These more general expansions are pertinent for describing radiation involving both incoming and outgoing states \cite{Chrusciel:1993hx}.}
    \be 
    \Phi(x)=\frac{\Sigma(u,\Omega)}{r^\Delta \ln^\alpha r}+\cdots\label{morefall}
    \ee where $\alpha\in\mathbb R$. Then the bulk-to-boundary correlator is determined by the limit 
    \be 
    D(u,\Omega;x')=\lim{}_+ r^\Delta \ln^\alpha r G(x;x').
    \ee It is straightforward to show that the Ward identities \eqref{Wardbtb} are still valid and thus the solution remains \eqref{bulktoboundary} which is independent of $\alpha$. Another possible extension would require $\Delta$ lies in the principal series of the M\"{o}bius group \footnote{In celestial holography, one commonly employs this choice of $\Delta$ \cite{Pasterski:2017kqt}. Note that it also arises in the  black hole perturbation theory because of the long-range gravitational potential \cite{1988sfbh.book.....F}.}, namely,  $\Delta=1+i\lambda$ with $\lambda\in\mathbb R$
    \be 
    \Phi(x)=\frac{\Sigma(u,\Omega)}{r^{1+i\lambda}}+\cdots+\text{h.c.}.
    \ee The form of the bulk-to-boundary correlator holds in this case as well. 
    \item \textbf{$i\epsilon$ prescription}. In our case, there are two ways to insert the $i\epsilon$. They are 
    \bea 
    D_F(u,\Omega;x')=\frac{C_s}{(u+n\cdot x'-i\epsilon)^\Delta},\quad D_{\bar F}(u,\Omega;x')=\frac{C_s}{(u+n\cdot x'+i\epsilon)^\Delta}.
    \eea When $\Delta=1$, $D_F(u,\Omega;x')$ is the same bulk-to-boundary propagator in the usual massless scalar field theory. This should be true for more general $\Delta$ via analytical  continuity. The second choice is not independent since it is the complex conjugate of $D_F$
    \be 
    D_{\bar F}(u,\Omega;x')=D^*_F(u,\Omega;x').
    \ee Therefore, $D_{\bar F}(u,\Omega;x')$ is related to the following quantity 
    \bea 
    D_{\bar F}(u,\Omega;x')=\langle \Phi(x')\Sigma(u,\Omega)\rangle.
    \eea This can be obtained by the following limit
    \be 
    D_{\bar F}(u,\Omega;x')=\lim{}_+r^\Delta G_{\bar F}(x;x')=\lim{}_+r^\Delta \langle \bar{\text{T}} \Phi(x)\Phi(x')\rangle.
    \ee The symbol $\bar{\text{T}}$ is the anti-time ordering operator that is defined as follows
    \be 
   \bar{ \text{T}}\Phi(x)\Phi(x')=\theta(x^0-x'^0)\Phi(x')\Phi(x)+\theta(x'^0-x^0)\Phi(x)\Phi(x').
    \ee 
    In classical linear response theory, the important quantity is the so-called retarded Green's function, which is proportional to the commutator of the bulk fields 
    \be 
    G_R(x;x')=i\theta(x^0-x'^0)\langle [\Phi(x),\Phi(x')]\rangle=i\theta(x^0-x'^0)\left(G_F(x;x')-G_{\bar F}(x;x')\right).
    \ee Extrapolating $\Phi(x)$ to the null boundary $\mathcal I^+$, then we may also define the retarded bulk-to-boundary correlator $D_R( u,\Omega;x')$ which is essentially the imaginary part of $D_F$
    \bea 
    D_R( u,\Omega;x')=i \left(D_F(u,\Omega;x')-D_{\bar F}(u,\Omega;x')\right)=-2\text{Im} D_F(u,\Omega;x').
    \eea When there is a classical source $J(x)$ coupled to the field $\Phi(x)$ in the bulk, the original action is deformed by a term 
    \be 
    \delta S=\mu\int d^4 x J(x)\Phi(x)
    \ee where $\mu$ is a small constant. Then at the leading order of $\mu$, the expectation value of the boundary field $\Sigma$ is universal 
    \bea 
   \Sigma_J(u,\Omega)=\mu \int d^4 x' J(x') D_R(u,\Omega;x').
    \eea We show the above equation in figure \ref{perturb}. 
    
    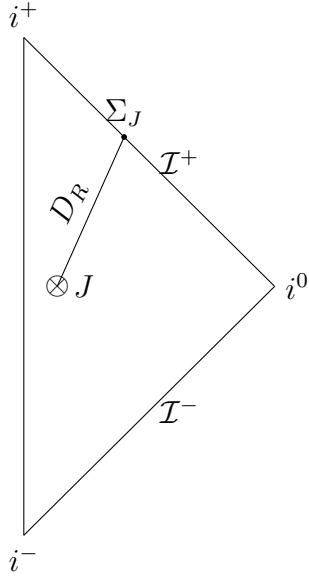
\begin{figure}
    \begin{center}
\begin{tikzpicture}[scale=1.1,>=stealth]

\coordinate (iplus) at (0,3);    
\coordinate (i0)   at (3,0);    
\coordinate (iminus) at (0,-3); 

\draw (iplus) -- (i0) node[midway,right] {$\mathcal{I}^+$};
\draw (i0) -- (iminus) node[midway,right] {$\mathcal{I}^-$};
\draw (iminus) -- (iplus);

\node[above] at (iplus) {$i^+$};
\node[right] at (i0)   {$i^0$};
\node[below] at (iminus) {$i^-$};

\node at (0.4,0) {$\otimes$};
\node[anchor=west, xshift=2pt] at (0.4,0) {$J$};

\coordinate (sigmaJ) at (1.2,1.8);
\filldraw (sigmaJ) circle (0.8pt) node[right,above] {$\Sigma_J$};
\draw (0.4,0) -- (sigmaJ) node[midway,above,sloped] {$D_R$};
\end{tikzpicture}
\caption{The classical value of the boundary field is corrected by the source in the bulk. We use the retarded bulk-to-boundary correlator since we are working in classical physics here. The leading order correction is universal.}\label{perturb}
\end{center}\end{figure}
   \item \textbf{Carrollian amplitude}. When perturbation theory is valid in the bulk, the bulk-to-boundary correlator is one of the key elements in Carrollian amplitude. Its Fourier transformation is 
   \bea 
   D(u,\Omega;x')=\frac{C_si^\Delta}{\Gamma(\Delta)}\int_0^\infty d\omega \omega^{\Delta-1}  e^{-i\omega(u+n\cdot x'-i\epsilon)}.
   \eea Therefore, the Carrollian amplitude is related to the momentum space connected Green's function through the modified Mellin transformation
   \bea 
   \langle \prod_{j=1}^m\Sigma_j(u_j,\Omega_j,\sigma_j)\rangle=\left(\prod_{j=1}^m \frac{C_{s,j} i^{\Delta_j}}{\Gamma(\Delta_j)}\right)\left(\prod_{j=1}^m\int_0^\infty d\omega_j \omega_j^{\Delta_j-1}e^{-i\sigma_j\omega_j{u_j}}\right) \mathcal G(p_1,\cdots,p_n).\label{Green}
   \eea The subscript $j$ is to distinguish operators inserted at different positions. This can be shown as follows \cite{Liu:2024nfc} 
   \bea 
  && \langle \prod_{j=1}^m \Sigma_j(u_j,\Omega_j,\sigma_j)\rangle\nn\\&=&\left(\prod_{j=1}^m \int d^4y_j D(u_j,\Omega_j;y_j)\right)G(y_1,y_2,\cdots,y_m)\nn\\&=&\left(\prod_{j=1}^m \int d^4 y_j\frac{C_{s,j}i^{\Delta_j}}{\Gamma(\Delta_j)}\int d\omega_j \omega_j^{\Delta_j-1}e^{-i\sigma_j\omega_j(u_j+n_j\cdot y_j)}\right)G(y_1,y_2,\cdots,y_m)\nn\\&=&\left(\prod_{j=1}^m \frac{C_{s,j} i^{\Delta_j}}{\Gamma(\Delta_j)}\right)\left(\prod_{j=1}^m\int_0^\infty d\omega_j \omega_j^{\Delta_j-1}e^{-i\sigma_j\omega_j{ u_j}}\right)  \mathcal G(p_1,p_2,\cdots,p_m)
   \eea where the Green's function $G(y_1,y_2,\cdots,y_m)$ is amputated and $\mathcal G(p_1,p_2,\cdots,p_m)$ is its Fourier transform in momentum space  
   \bea 
   \mathcal G(p_1,p_2,\cdots,p_m)=\prod_{j=1}^m \int d^4y_j e^{-ip_j\cdot y_j}G(y_1,y_2,\cdots,y_m)
   \eea where $p_j^\mu=\sigma_j \omega_j n^\mu_j$ and thus the momentum is on-shell.
   The derivation does not depend on the canonical quantization and the concept of particle. Certainly it depends on the existence of asymptotic states on future and past null infinity. To relate it to the scattering amplitude, we should use the LSZ reduction formula 
   \cite{Lehmann:1954rq} which states that the scattering amplitude $\mathcal A$ is proportional to the amputated Green's function up to the external lines. The momentum in the amputated Green's function should be on-shell. Indeed, we choose $\Delta=1$ and rewrite the correlator as \footnote{Here we have been slightly sloppy. As will be discussed in the final section, the normalization constant $C_{s,j}$ is given by $C_{s,j} \propto {Z}_j$ when the vacuum is stable. The field $\Sigma$ should be renormalized as $\Sigma_r = Z^{-1/2}\Sigma$. The remaining factor $\sqrt{Z}$ is then absorbed into the scattering amplitude $\mathcal{A}$ \cite{1995iqft.book.....P}, leaving an overall factor of $1/(8\pi^2)$.}
   \bea 
\langle \prod_{j=1}^m \Sigma_j(u_j,\Omega_j,\sigma_j)\rangle=\left( \frac{i}{8\pi^2}\right)^m\left(\prod_{j=1}^m \int_0^\infty d\omega_j e^{-i\sigma_j\omega_j u_j}\right) \mathcal A(p_1,p_2,\cdots,p_m).
   \eea 
\end{itemize}

\subsection{Spinor}
In this subsection, we will extend the previous discussion to fermion operators. 
\subsubsection{Conventions}
The Lorentz transformation law of a Dirac spinor $\Psi(x)$ is 
\be 
\Psi'(x')=\Lambda_{1/2}\Psi(x)
\ee where 
\be 
\Lambda_{1/2}=e^{-\frac{i}{2}\omega_{\mu\nu}S^{\mu\nu}}.
\ee 
The inverse of $\Lambda_{1/2}$ will be denoted as $\Lambda_{-1/2}$.
The antisymmetric generator $S^{\mu\nu}$ is 
\be 
S^{\mu\nu}=\frac{i}{4}[\gamma^\mu,\gamma^\nu].
\ee The gamma matrices are chosen as 
\bea 
\gamma^\mu=\left(\begin{array}{cc}0&\sigma^\mu\\\bar\sigma^\mu&0\end{array}\right)
\eea with 
\bea 
\sigma^\mu_{a\dot a}=(1,\sigma^i),\quad \bar\sigma^{\mu\dot a a}=(1,-\sigma^i).
\eea
The lowercase Latin letters $a,b,\cdots$ denotes the left-hand Weyl spinor indices while the dotted letters $\dot a,\dot b,\cdots$ represents the right hand Weyl spinor indices.
The three Pauli matrices $\sigma^i$ are 
\bea 
\sigma^1=\left(\begin{array}{cc}0&1\\1&0\end{array}\right),\quad \sigma^2=\left(\begin{array}{cc}0&-i\\ i&0\end{array}\right),\quad \sigma^3=\left(\begin{array}{cc}1&0\\0&-1\end{array}\right)
\eea such that the $\gamma^\mu$ matrices obey the anti-commuting relations
\be 
\{\gamma^\mu,\gamma^\nu\}=-2\eta^{\mu\nu}.
\ee 
Note that the $4\times 4$ matrix $\Lambda_{1/2}$ can be decomposed as the direct sum of the left-hand and right-hand part
\be 
\Lambda_{1/2}=\left(\begin{array}{cc}\Lambda_{1/2}^L&0\\ 0&\Lambda_{1/2}^R\end{array}\right).
\ee Both of $\Lambda_{1/2}^L$ and $\Lambda_{1/2}^R$ are $2\times 2$ matrices.
A useful commutator is
\be
    [S^{\mu\nu},\gamma^\rho]=i\left(\eta^{\mu\rho}\gamma^\nu-\eta^{\nu\rho}\gamma^\mu\right).
\ee  
One can also define a $\gamma^5$ matrix 
\be
\gamma^5=-i\gamma^0\gamma^1\gamma^2\gamma^3=\left(\begin{array}{cc}-1&0\\ 0&1\end{array}\right)
\ee that is anticommute with all other $\gamma^\mu$ matrices
\be 
\left\{\gamma^5,\gamma^\mu\right\}=0.
\ee 

\subsubsection{Bulk-to-boundary correlators}
Similar to the scalar case, we focus on the two-point correlator 
\be 
G(x;y)=\langle \text{T} \Psi(x)\bar \Psi(y)\rangle
\ee where 
\bea \bar\Psi(y)=\Psi^\dagger(y)\gamma^0.
\eea
The time-ordering of a fermionic field is 
\be 
\text{T}\Psi(x)\bar\Psi(y)=\theta(x^0-y^0)\Psi(x)\bar\Psi(y)-\theta(y^0-x^0)\bar\Psi(y)\Psi(x).
\ee Note that there is a minus sign for the exchange of the fermionic field. The Ward identities from Poincar\'e invariance are 
\bs\begin{align}
    \partial_\mu^x G(x;y)+\partial_\mu^y G(x;y)&=0,\\
    x_{[\nu}\partial^x_{\mu]}G(x;y)+y_{[\nu}\partial^y_{\mu]}G(x;y)&=\frac{i}{2}[S_{\mu\nu},G(x;y)].
\end{align}\es The two-point correlator $G(x;y)$ is a $4\times 4$ matrix since the fermion fields contain spinor indices. The right hand side of the second line is understood as the commutator between two matrices $S_{\mu\nu}$ and $G(x;y)$. The fermionic operator $\Psi(x)$ may be expanded asymptotically 
\bea 
\Psi(x)=\left\{\begin{array}{cc} \frac{\psi(u,\Omega)}{r^{\Delta}}+\cdots,& \text{near}\  \mathcal I^+,\\
\frac{\psi^{(-)}(v,\Omega)}{r^{\Delta}}+\cdots,&\text{near}\  \mathcal{I}^-.\end{array}\right.\label{falloffpp}
\eea where $\psi(u,\Omega)$ and $\psi^{(-)}(v,\Omega)$ are the boundary field on $\mathcal I^+$ and $\mathcal I^-$, respectively. Now the bulk-to-boundary correlators are defined as 
\bs\begin{align}
    D(u,\Omega;x')&=\langle \psi(u,\Omega)\bar\Psi(x')\rangle,\\
    D^{(-)}(v,\Omega;x')&=-\langle
    \bar \Psi(x')\psi(v,\Omega)\rangle.
\end{align}\es One can also extrapolate the field $\bar\Psi(x)$ to the boundary to obtain two more correlators. We will focus on $D(u,\Omega;x')$ since the discussion on the other correlators are parallel. We may expand $G(x;x')$ near $\mathcal I^+$ to obtain 
\bea 
G(x;x')=\frac{D(u,\Omega;x')}{r^{\Delta}}+\cdots.\label{fallofffer}
\eea Then the Ward identities for $D(u,\Omega;x')$ are
\bs\begin{align}
    n_\mu \partial_u D(u,\Omega;x')&=\partial'_\mu D(u,\Omega;x'),\label{fermionD}\\
    x'^{[\nu}\partial'^{\mu]}D(u,\Omega;x')&=\left(-\frac{1}{2}n^{\mu\nu}u\partial_u-\frac{\Delta}{2}n^{\mu\nu}+\frac{1}{2}Y^{\mu\nu A}\partial_A\right)D(u,\Omega;x')+\frac{i}{2}[S^{\mu\nu},D(u,\Omega;x')].\label{fermionD2}
\end{align}\es Solving \eqref{fermionD} we still find 
\be 
D(u,\Omega;x')=F(u+n\cdot x',\Omega)
\ee where $F$ is a $4\times 4$ matrix valued function of $u+n\cdot x$ and $\Omega$. The $4\times 4$ matrix can be expanded using the basis 
\be 
1,\gamma^\mu,S^{\mu\nu},\gamma^5\gamma^\mu,\gamma^5.\label{basisF}
\ee Namely, 
\be 
F=f_1(u+n\cdot x',\Omega)+f^\mu_2(u+n\cdot x',\Omega)\gamma_\mu+f_3^{\mu\nu}(u+n\cdot x',\Omega) S_{\mu \nu}+f_4^\mu(u+n\cdot x',\Omega)\gamma^5\gamma_\mu+f_5(u+n\cdot x',\Omega)\gamma^5.\label{exF}
\ee Unfortunately, there would be 16 independent functions to be determined. To simplify the derivation, we notice that $\Omega$ is in one-to-one correspondence with $n^\mu$. Therefore, we may rewrite $F$ as 
\be 
F=F(u+n\cdot x',n).
\ee Each $\gamma^\mu$ should be paired with a vector when expanding $F$ in the basis \eqref{basisF}. Therefore, the unique combination is  $n_\mu\gamma^\mu=\slashed n$. The expansion \eqref{exF} is simplified to 
\bea 
F=f_1(u+n\cdot x',\Omega)+f_2(u+n\cdot x',\Omega)\slashed n+f_4^\mu(u+n\cdot x',\Omega)\gamma^5\slashed n+f_5(u+n\cdot x',\Omega)\gamma^5.\label{exF2}
\eea The functions $f_3^{\mu\nu}$ disappear since $S^{\mu\nu}n_\mu n_\nu=0$. To justify the expansion \eqref{exF2}, we notice that the bulk-to-bulk correlator $G(x;x')$ should be a function of $(x-x')^\mu=r n^\mu+u\bar m^\mu-x'^\mu$. When expanding $G(x;x')$ in the basis, one should pair $(x-x')^\mu$ and $\gamma_\mu$
\be 
(x-x')_\mu \gamma^\mu=r \slashed n+u \slashed{\bar m}-\slashed x'.
\ee At the leading order, only $\slashed n$ contributes to the bulk-to-boundary correlator. Substituting \eqref{exF2} into \eqref{fermionD2}, we find 
\bea 
0&=&-\frac{1}{2}n^{\mu\nu}\left[(u+n\cdot x)f_1'+\Delta f_1\right]-\frac{1}{2}n^{\mu\nu}\left[(u+n\cdot x)f_2'+(\Delta+1) f_2\right]\slashed n \nn\\&&-\frac{1}{2}n^{\mu\nu}\left[(u+n\cdot x)f_4'+(\Delta +1)f_4\right]\gamma^5\slashed n-\frac{1}{2}n^{\mu\nu}\left[(u+n\cdot x)f_5'+\Delta f_5\right]\gamma^5\nn\\&&+n^{[\nu}Y^{\mu]A}\delta_A f_1+n^{[\nu}Y^{\mu]A}\slashed n\delta_A f_2+n^{[\nu}Y^{\mu]A}\gamma^5\slashed n\delta_A f_4+n^{[\nu}Y^{\mu]A}\delta_A f_5\gamma^5.
\eea Similar to the scalar case, multiplying both sides with $Y_\mu^B$ and then 
\be 
\delta_Af_1=\delta_Af_2=\delta_Af_4=\delta_Af_5=0
\ee due to the basis $1,\gamma^\mu,\gamma^5\gamma^\mu,\gamma^5$ are independent. Now we can solve the Ward identities by 
\bs\begin{align}
    f_1&=\frac{C_1}{(u+n\cdot x')^\Delta},\\
    f_2&=\frac{C_2}{(u+n\cdot x')^{\Delta+1}},\\
    f_4&=\frac{C_4}{(u+n\cdot x')^{\Delta+1}},\\
    f_5&=\frac{C_5}{(u+n\cdot x')^{\Delta}}.
\end{align}\es where $C_1,C_2,C_4,C_5$ are independent constants. Therefore, the bulk-to-boundary correlator is 
\bea 
D(u,\Omega;x')=\frac{C_s}{(u+n\cdot x'-i\epsilon)^\Delta}+\frac{C_f\slashed n}{(u+n\cdot x'-i\epsilon)^{\Delta+1}}+\frac{C_{pf}\gamma^5\slashed n}{(u+n\cdot x'-i\epsilon)^{\Delta+1}}+\frac{C_{ps}\gamma^5}{(u+n\cdot x'-i\epsilon)^{\Delta}}.
\eea We have inserted $i\epsilon$ into the correlator and relabeled the coefficients with suitable subscripts. There are four independent branches for the fermion bulk-to-boundary correlator $D(u,\Omega;x')$. The first branch is exactly the same as the scalar case, therefore we may call it  scalar branch and denote the normalization coefficient as $C_s$. The second branch is new compared to the scalar case. Therefore, we may call it  fermionic branch and denote the normalization coefficient as $C_f$. The third branch contains an extra $\gamma^5$ factor compared with the fermionic branch. We may call it pseudo-fermionic branch and the normalization coefficient is $C_{pf}$. Similarly, the last branch is called  peudo-scalar branch and the normalization factor is $C_{ps}$. Note that the fall-off of the fermionic branch transforms as 
\be 
\frac{1}{\widehat u^{\Delta+1}}\to \frac{\Gamma^{\Delta+1}}{\widehat u^{\Delta+1}}
\ee under Lorentz rotations. The scaling dimension $\Delta+1$ does not match with the fall-off exponent $\Delta$ for the fermions. This phenomenon would affect the exponents in the Carrollian amplitude,a s will be discussed in the next section. Another notable thing is that the pseudo-scalar and pseudo-fermionic branches can not appear in a parity invariant theory since $\gamma^5$ breaks it. 
Therefore, we will only consider the scalar and fermionic branch in the bulk-to-boundary correlator
\be 
D(u,\Omega;x')=\frac{C_s}{(u+n\cdot x'-i\epsilon)^\Delta}+\frac{C_f\slashed n}{(u+n\cdot x'-i\epsilon)^{\Delta+1}}.\label{parityinv}
\ee 
\subsubsection{Dirac fields}
Now we will use free Dirac theory to illustrate the fermionic branch. In principle, we can begin with the propagator 
\be 
G(x;x')=-\frac{i}{2\pi^2}\frac{\gamma^\mu (x-x')_\mu}{((x-x')^2+i\epsilon)^2}
\ee and then reduces $\Psi(x)$ (or $\bar\Psi(x')$) to the boundary. Then the bulk-to-boundary propagators are 
\bs\label{bulktoboudnr}\begin{align}
    \langle \psi(u,\Omega)\bar\Psi(x')\rangle&=\lim{}_+r G(x;x')=-\frac{i \slashed n}{8\pi^2(u+n\cdot x'-i\epsilon)^2},\label{psibarPsi}\\
    \langle \bar\psi(u,\Omega)\Psi(x')\rangle&=-\lim{}_+r G(x';x)=-\frac{i\slashed n}{8\pi^2(u+n\cdot x'-i\epsilon)^2},\label{psibarPsi2}\\\langle \bar \Psi(x')\psi^{(-)}(v,\Omega)\rangle&=-\lim{}_{-}r G(x;x')=\frac{i\slashed {\bar n}}{8\pi^2(v-\bar n\cdot x'+i\epsilon)^2},\\
    \langle \Psi(x')\bar \psi^{(-)}(v,\Omega)\rangle&=\lim{}_-r G(x';x)=\frac{i{\slashed{\bar n}}}{8\pi^2(v-\bar n\cdot x'+i\epsilon)^2}.
\end{align}
\es The propagator \eqref{psibarPsi} matches with the fermionic branch with 
\be 
C_f=-\frac{i}{8\pi^2}.
\ee There is no scalar branch in free Dirac theory.

\section{Carrollian amplitude for fermionic theory}\label{cmpl}
With the explicity{explicit} bulk-to-boundary propagators, one can  already use the position space Feynman rule to construct the Carrollian amplitude, as has been done in the scalar theory \cite{Liu:2024nfc}. However, the connection to the momentum space scattering amplitude is unclear in this case, especially when one uses modern technology such as spinor helicity formalism \cite{Mason:2023mti}.
In a massless fermionic theory such as QED, the Fourier transform of the bulk-to-boundary propagator is 
\bea 
D(u,\Omega;x')=-\frac{i\slashed n}{8\pi^2(u+n\cdot x'-i\epsilon)^2}=\frac{i\slashed n}{8\pi^2}\int_0^\infty d\omega \omega e^{-i\omega(u+n\cdot x'-i\epsilon)}.\label{bulktoboundaryfermoin}
\eea 
Note that there is a factor $\omega$ that is inserted into the integral, in contrast with the scalar theory. In \cite{Donnay:2022wvx,Salzer:2023jqv}, the momentum space amplitude is related to the Carrollian amplitude via the Fourier transform 
\bea &&\langle \prod_{j=1}^{m}\Sigma_j(u_j,\Omega_j,\sigma_j)\rangle=(\frac{i}{8\pi^2})^{m}\prod_{j=1}^{m} \int d\omega_j e^{-i\sigma_j\omega_j u_j}\mathcal{A}(\bm p_1,\bm p_2,\cdots,\bm p_{m})\nn\\\label{Carrollianamp}
\eea where $\Sigma_j$ denotes a scalar operator (or a spin $s$ bosonic operator) inserted at the boundary with position $(u_j,\Omega_j)$ and  the symbol $\sigma_j, j=1,2,\cdots,m$ is designed to distinguish the outgoing and incoming states via the convention
\bea 
\sigma_j=\left\{\begin{array}{cl}+1&\text{outgoing state},\\
-1&\text{incoming state}.\end{array}\right.
\eea The notation $\mathcal A$ denotes the scattering amplitude 
of $m$ particles whose momentums are
\be 
p_j^\mu=(\omega_j,\bm p_j)=\sigma_j\omega_j(1, \bm n_j)=\sigma_j \omega_j n_j^\mu,\quad j=1,2,\cdots,m.
\ee 
Note that there is no $\omega$ in the integral transform.
Modified Mellin transform can also be used to obtain the correlators 
\bea 
\langle \prod_{j=1}^m \Sigma_j^{(k_j)}(u_j,\Omega_j,\sigma_j)\rangle=(\frac{i}{8\pi^2 })^{m}\prod_{j=1}^{m} \int d\omega_j (-i\sigma_j\omega_j)^{k_j}e^{-i\sigma_j\omega_j u_j}\mathcal{A}(\bm p_1,\bm p_2,\cdots,\bm p_{m})
\eea where 
\be 
\Sigma^{(k)}(u,\Omega)=\left(\frac{d}{du}\right)^{k}\Sigma(u,\Omega),\quad k=0,1,2,\cdots.
\ee The natural numbers $k_j$ may be analytically continued to the real or complex values away from singularities. Note that when $k$ is not an integer, the operator $\Sigma^{(k)}$ would be non-local in the sense of Riemann-Liouville fractional integral \cite{2006Theory}. In the following, we will show that the most natural choices of $k_j$ are half-integers for Carrollian correlators of fermionic operators, i.e.
\be 
k_j=\frac{1}{2},\ \frac{3}{2},\ \frac{5}{2},\ \cdots.
\ee 
 \subsection{Canonical quantization}
 We will adopt the canoincal quantization to discuss the details.
A Dirac field can be written as the mode expansion
\bea
    \Psi(x) = \sum_{s=\pm} \int \frac{d^3 \bm p}{(2\pi)^3 2\omega}[b_{s,\bm p}u_s(\bm p)e^{ipx}+d_{s,\bm p}^{\dagger}v_s(\bm p )e^{-ipx}]\label{psi21}
\eea
where $u_{s}(\bm p)$ and $v_{s}(\bm p)$ satisfy the equation of motion 
\bea
\slashed p u_{s}(\bm p) =\slashed p v_{s}(\bm p)=0.\label{uvs}
\eea The operator $d_{s,\bm p}^\dagger$ creates an anti-fermion particle with momentum $\bm p$ and helicity $s$. On the other hand, the operator $b_{s,\bm p}^\dagger$  creates a fermion particle with momentum $\bm p$. For Majorana fermions, we have the condition 
\be 
d_s(\bm p)=b_s(\bm p).
\ee

The four-momentum $p_{\mu}$ is null and we can factorize it as 
\bea
p_{\mu}=\omega n_{\mu},
\eea where $n_\mu$ is the null vector that is determined by the spatial direction of the momentum. A frequently used trick is to switch the null vectors to the matrix form 
\bea
p_{a\dot a}=p_{\mu}\sigma^{\mu}_{a \dot a}=\omega n_{a\dot a}=-\omega\lambda_a \lambda^{\dagger}_{\dot a}\label{padota}
\eea where 
\be 
n_{a\dot a}=n_\mu\sigma^\mu_{a\dot a}
\ee and the two-component Weyl spinor $\lambda_a$ is \be 
\lambda_a =i\sqrt{\frac{2}{1+|z|^2}}(1,-z).\label{deflambda}\ee Its Hermitian conjugate is 
\be 
\lambda_{\dot a}^\dagger=\left(\lambda_a\right)^{\dagger}= -i\sqrt{\frac{2}{1+|z|^2}}(1,-\bar z).
\ee 
One can also find a dual Weyl spinor $\kappa_a$ 
\be 
\kappa_a=i\sqrt{\frac{2}{1+|z|^2}}(\bar z,1)
\ee whose product with $\lambda_a$ is normalized to 2.
\be 
\lambda^a\kappa_a=2.
\ee The spinors $\lambda_a,\ \kappa_a$ commutes with each other and these commuting spinors  are called twistors in \cite{2007qft..book.....S}.
The solutions of $\eqref{uvs}$ are 
\bea
u_{+}(\bm p)=v_{-}(\bm p)=\sqrt{\omega}\left(\begin{array}{c}
0\\
\lambda^{\dagger \dot a}
\end{array}\right),\quad u_{-}(\bm p)=v_{+}(\bm p)=\sqrt{\omega}\left(\begin{array}{c}
\lambda_a\\
0
\end{array}\right).\label{slolutionu}
\eea
The annihilation and creation operations $b_{s,\bm p},d^{\dagger}_{s,\bm p}$ satisfy the anticommutators
\bea
 \left\{b_{ s,\bm p},b^{\dagger}_{s',\bm p'} \right\}=\left\{d_{s,\bm p},d^{\dagger}_{s',\bm p'}\right\}=(2\pi)^32\omega\delta_{ss'}\delta(\bm p-\bm p'),
\eea
where the vanishing anticommutators are omitted.
From the fall-off condition near $\mathcal I^{\pm}$, we can expand the plane wave as the superposition of the spherical waves and read out the mode expansion of the boundary fields. 
To be more precise, we expand the Dirac field near $\mathcal I^+$ as \begin{subequations}
\begin{align} 
\Psi&=\frac{1}{r}\psi(u,\Omega)+\frac{1}{r^2}\psi^{(2)}(u,\Omega)+\mathcal{O}(r^{-3}),\label{psiu}\\ \bar\Psi&=\frac{1}{r}\bar\psi(u,\Omega)+\frac{1}{r^2}\bar\psi^{(2)}(u,\Omega)+\mathcal{O}(r^{-3})
\end{align}\end{subequations} where 
\bea 
\psi(u,\Omega)=\left(\begin{array}{c}\lambda_a F(u,\Omega)\\ \lambda^{\dagger \dot a}\bar G(u,\Omega)\end{array}\right),\quad \bar \psi(u,\Omega)=(\lambda^a G(u,\Omega),\lambda^\dagger_{\dot a}\overline F(u,\Omega)).\label{defpsi}
\eea 
The $\bar F$ denotes the complex conjugate of $F$. Similar conventions are used for the field $G$. A general Dirac spinor would be in the form of \be 
\Psi=\left(\begin{array}{c}\chi_a\\ \xi^{\dagger \dot a}\end{array}\right)
\ee and the left-hand Weyl spinor $\chi_a$ may be expanded in the basis of $\lambda_a$ and $\kappa_a$. However, at the leading order of the asymptotic expansion \eqref{psiu}, the left-hand part of the field $\psi(u,\Omega)$ is solved to be proportional to $\lambda_a$. Similarly, we find the expansion near $\mathcal I^-$
\begin{subequations}
\begin{align} 
\Psi&=\frac{1}{r}\psi^{(-)}(v,\Omega)+\frac{1}{r^2}\psi^{(2)(-)}(v,\Omega)+\mathcal{O}(r^{-3}),\\ \bar\Psi&=\frac{1}{r}\bar\psi^{(-)}(v,\Omega)+\frac{1}{r^2}\bar\psi^{(2)(-)}(v,\Omega)+\mathcal{O}(r^{-3})
\end{align}\end{subequations} with
\bea 
\psi^{(-)}( v,\Omega)=\left(\begin{array}{c}\lambda_a(\Omega^{\text P}) F^{(-)}(v,\Omega^{\text P})\\ \lambda^{\dagger \dot a}(\Omega^{\text P})\bar G^{(-)}(v,\Omega^{\text P})\end{array}\right)\label{psim}\eea and \bea \bar \psi^{(-)}(v,\Omega)=(\lambda^a(\Omega^{\text P}) G^{(-)}(v,\Omega^{\text P}),\lambda_{\dot a}^\dagger(\Omega^{\text P}) \bar F^{(-)}(v,\Omega^{\text P})).\label{barpsim}
\eea  The notation $\Omega^{\text{P}}$ represents the antipodal point of $\Omega=(\theta,\phi)$,
\be 
\Omega^{\text P}=(\pi-\theta,\pi+\phi).
\ee Therefore, the notation $\bm p^{\text P}$ denotes the antipodal map of $\bm p$, in spherical coordinates, this is 
\be 
\bm p^{\text P}=(\omega,\pi-\theta,\pi+\phi).
\ee In this notation, 
the  mode expansions of the boundary fields are 
\bs\label{qc}\begin{align}
 F(u, \Omega)=-\frac{i}{8 \pi^{2}} \int_{0}^{\infty} d \omega \sqrt{\omega}\left(b_{-,\bm p} e^{-i \omega u}-d_{+,\bm p}^{\dagger} e^{i \omega u}\right),\\
 F^{ (-)}(v, \Omega)=\frac{i}{8 \pi^{2}} \int_{0}^{\infty} d \omega \sqrt{\omega}\left(b_{-,\bm p} e^{-i \omega v}-d_{+, \bm p}^{\dagger} e^{i \omega v}\right),\\
 \bar{G}(u, \Omega)=-\frac{i}{8 \pi^{2}} \int_{0}^{\infty} d \omega \sqrt{\omega}\left(b_{+,\bm p} e^{-i \omega u}-d_{-,\bm p}^{\dagger} e^{i \omega u}\right) ,\\
 \bar{G}^{(-)}(v, \Omega)=\frac{i}{8 \pi^{2}} \int_{0}^{\infty} d \omega \sqrt{\omega}\left(b_{+, \bm p} e^{-i \omega v}-d_{-, \bm p}^{\dagger} e^{i \omega v}\right) .
\end{align}\es 
An asymptotic outgoing state may be defined as 
\be 
|F(u,\Omega)\rangle=F(u,\Omega)|0\rangle=\frac{i}{8\pi^2}\int_0^\infty d\omega \sqrt{\omega}e^{i\omega u}|p,+,+\rangle
\ee where $|p,+,+\rangle$ is an outgoing anti-fermion particle state with definite momentum $\bm p$ and positive helicity
\be 
|\bm p,+,+\rangle=d_{+,\bm p}^\dagger |0\rangle.
\ee 
We used the notation $|\bm p,s,a\rangle$ to represents a state with momentum $\bm p$, helicity $s$. The label $a=\pm$ is to distinguish the fermion and antifermion. More precisely, 
\bea 
a=\left\{\begin{array}{cc}+&\ \text{anti-fermion}\\ -&\ \text{fermion}.\end{array}\right.
\eea 
Therefore, the state $|F(u,\Omega)\rangle$ represents an outgoing anti-fermion with positive helicity and  is located at $(u,\Omega)$.
 Acting on the vacuum from the right, we find 
 \be 
 \langle F(u,\Omega)|=\langle 0|F(u,\Omega)=-\frac{i}{8\pi^2}\int_0^\infty d\omega \sqrt{\omega}e^{-i\omega u}\langle \bm p,-,-|
 \ee which is an outgoing fermion with negative helicity and is located at $(u,\Omega)$. Note that this is not the Hermitian conjugate of $|F(u,\Omega)\rangle$ in our notation. On the other hand, we find 
 \be 
\left( |F(u,\Omega)\rangle \right)^\dagger=\langle 0|\overline F(u,\Omega)=\langle \overline F(u,\Omega)|=-\frac{i}{8\pi^2}\int_0^\infty d\omega \sqrt{\omega} e^{-i\omega u}\langle \bm p,+,+|.
 \ee In the table \ref{table1}, we summarize the properties of the states  that are constructed from mode expansions \eqref{qc}. In this table, outgoing (incoming) states are represented as bra $\langle out|$ (ket $|in\rangle$) states. 
 \begin{table}
\centering 
\renewcommand{\arraystretch}{1.5}
\begin{tabular}{|c|c|c|c|c|c|c|c|c|}\hline
State&$\langle F|$&$\langle G|$&$\langle \bar F|$&$\langle \bar G|$&$|F^{(-)}\rangle$&$| G^{(-)}\rangle$&$|\bar F^{(-)}\rangle$&$|\bar  G^{(-)}\rangle$\\\hline
$\sigma$&$+$&$+$&$+$&$+$&$-$&$-$&$-$&$-$\\\hline
$s$&$-$&$-$&$+$&$+$&$+$&$+$&$-$&$-$\\\hline
$a$&$-$&$+$&$+$&$-$&$+$&$-$&$-$&$+$\\\hline
\end{tabular}
\caption{\centering{States at the null boundaries}}
\label{table1}
\end{table}
Motivated by the crossing symmetry of the scattering amplitude, we may flip the sign of the helicity and exchange the fermion and anti-fermion for the incoming states and then define the fields
 \bs\label{antipodal}\begin{align}
     F(u,\Omega,-)&=-F^{(-)}(v,\Omega)\Big|_{b_s\leftrightarrow d_{-s}, v\to -u}=-\frac{i}{8\pi^2}\int_0^\infty d\omega \sqrt{\omega}\left(d_{+,\bm p}e^{i\omega u}-b_{-,\bm p}^\dagger e^{-i\omega u}\right),\\
     \bar G(u,\Omega,-)&=-\bar G^{(-)}(v,\Omega)\Big|_{b_s\leftrightarrow d_{-s}, v\to -u}=-\frac{i}{8\pi^2}\int_0^\infty d\omega \sqrt{\omega}\left(d_{-,\bm p}e^{i\omega u}-b_{+,\bm p}^\dagger e^{-i\omega u}\right).
 \end{align}\es 
 The replacement $v\to -u$ may be confusing at first glance. However, one should note that crossing symmetry requires us to flip the sign of $p$ to $-p$ at the same time. 
 In spherical coordinates, this is equivalent to the following transformation 
 \be 
 (\omega,\bm p)\to (-\omega, \bm p^{\text P}).\label{flipp}
 \ee The sign flipping of $\omega$ is established by setting $v\to -u$ for the fields on $\mathcal I^-$. The coordinate $v$ becomes coordinate $u$ since the corresponding state becomes outgoing one.
 The transformation \eqref{flipp} also indicates us to send $\bm p$ to its antipodal point. However, this step has already been completed in \eqref{psim} where the spherical coordinates on the right-hand side are already $\Omega^{\text P}$. We have also added a minus sign before $F^{(-)}$ (and $G^{(-)}$) in the definition \eqref{antipodal}. This is a choice of convention. Now we can define the operator $\uppsi_s(u,\Omega,\sigma,a)$ as follows
\begin{align}
     \uppsi_s(u,\Omega,\sigma,a)=\left\{\begin{array}{cc} F(u,\Omega)& s=-,\ \sigma=+,\ a=-\\ \bar G(u,\Omega)&s=+,\ \sigma=+,\ a=-\\ 
     \bar F(u,\Omega)&s=+,\ \sigma=+,\ a=+\\ G(u,\Omega)&s=-,\ \sigma=+,\ a=+\\ F(u,\Omega,-)&s=+,\ \sigma=-,\ a=+\\
     \bar G(u,\Omega,-)&s=-,\ \sigma=-,\ a=+\\ \bar F(u,\Omega,-)& s=-,\ \sigma=-,\ a=-\\ G(u,\Omega,-)& s=+,\ \sigma=-,\ a=-.\end{array}\right.
 \end{align} Now we can define the Carrollian amplitude via the scattering amplitude for the fermionic operators
 \bea 
 \langle \prod_{j=1}^m \uppsi_{s_j}(u_j,\Omega_j,\sigma_j,a_j)\rangle=\left(-\frac{i}{8\pi^2}\right)^m \left(\prod_{j=1}^m \int_0^\infty d\omega_j \sqrt{\omega_j} e^{-i\sigma_j\omega_j u_j}\right)\mathcal A_{s_1s_2\cdots s_m}(1,2,\cdots,m)\label{caf}\nn\\
 \eea 
 where $\mathcal A$ is the $m$-point scattering amplitude in momentum space
 \be 
  \mathcal A_{s_1s_2\cdots s_m}(1,2,\cdots,m)=\langle \bm p_1,s_1,a_1,\sigma_1;\bm p_2,s_2,a_2,\sigma_2;\cdots;\bm p_m,s_m,a_m,\sigma_m\rangle.
 \ee It has already been adapted to the spinor helicity formalism. Interestingly, there is a factor of $\sqrt{\omega}$ in the Fourier transform, in contrast to the bosonic case (see eqn. \eqref{Carrollianamp}). Taking time derivatives, we can also find 
 \be 
  \langle \prod_{j=1}^m \uppsi^{(k_j)}_{s_j}(u_j,\Omega_j,\sigma_j,a_j)\rangle=\left(-\frac{i}{8\pi^2}\right)^m \left(\prod_{j=1}^m \int_0^\infty d\omega_j (-i\sigma_j\omega_j)^{k_j}\sqrt{\omega_j} e^{-i\sigma_j\omega_j u_j}\right)\mathcal A_{s_1s_2\cdots s_m}(1,2,\cdots,m).
 \ee Note that our result is consistent with the integral representation of the bulk-to-boundary propagator \eqref{bulktoboundaryfermoin}. As an illustration, we can make use of the mode expansions \eqref{qc} and \eqref{psi21} and then obtain
 \begin{align}
     \langle F(u,\Omega)\bar \Psi(x')\rangle&=\frac{i(0, \lambda_{\dot a}^\dagger)}{8\pi^2(u+n\cdot x'-i\epsilon)^2},\quad \langle \bar G(u,\Omega)\bar \Psi(x')\rangle=\frac{i(\lambda^a,0)}{8\pi^2(u+n\cdot x'-i\epsilon)^2}.
 \end{align}

 Substituting into \eqref{defpsi}, we reproduce the propagator \eqref{bulktoboundaryfermoin}. In the calculation, we used the identities
 \bea 
 n_{a\dot a}=n_\mu\sigma^\mu_{a\dot a}=-\lambda_a\lambda^\dagger_{\dot a},\quad n^{\dot a a}=n_\mu \bar\sigma^{\mu \dot a a}=-\lambda^{\dagger\dot a}\lambda^a.
 \eea 
When computing the Carrollian amplitude via position space Feynman rules, the factor $\sqrt{\omega}$ in the integral representation \eqref{bulktoboundaryfermoin} is absorbed into the momentum space amplitude, as seen in \eqref{slolutionu}. Consequently, a factor of $\sqrt{\omega}$ remains in the final Carrollian amplitude \eqref{caf}. This can also be understood from the relation between the Carrollian correlator and the amputated Green's function. For fermionic operators, this relation holds provided one replaces the bulk-to-boundary propagator with \eqref{bulktoboundaryfermoin}. Applying the LSZ reduction formula to connect the amputated Green's function to the scattering amplitude requires incorporating the external spinor lines. These external lines for Dirac fields, given in \eqref{slolutionu}, are each proportional to $\sqrt{\omega}$ for incoming and outgoing states, which directly explains the surviving factor in \eqref{caf}.
\\

 \textbf{Remarks.} 
 \begin{itemize}
     \item \textbf{Bulk-to-boundary propagators.} There are more bulk-to-boundary propagators can be found. We collect them in the following:
 \bs\begin{align}
 \langle F(u,\Omega)\bar \Psi(x')\rangle&=\frac{i(0, \lambda_{\dot a}^\dagger)}{8\pi^2(u+n\cdot x'-i\epsilon)^2},\quad \langle \bar  G(u,\Omega)\bar \Psi(x')\rangle=\frac{i(\lambda^a,0)}{8\pi^2(u+n\cdot x'-i\epsilon)^2},\\
     \langle G(u,\Omega)\Psi(x')\rangle&=\frac{i\left(\begin{array}{c}0\\ \lambda^{\dagger\dot a}\end{array}\right)}{8\pi^2(u+n\cdot x'-i\epsilon)^2},\quad \langle \bar F(u,\Omega)\Psi(x')\rangle=\frac{i\left(\begin{array}{c}\lambda_a\\0\end{array}\right)}{8\pi^2(u+n\cdot x'-i\epsilon)^2},\\
\langle \bar\Psi(x')F^{(-)}(v,\Omega)\rangle&=\frac{i(0,\lambda_{\dot a}^\dagger)}{8\pi^2(v -\bar n\cdot x'+i\epsilon)^2},\quad \langle \bar \Psi(x')\bar G^{(-)}(v,\Omega)\rangle=\frac{i(\lambda^a,0)}{8\pi^2(v-\bar n\cdot x'+i\epsilon)^2},\\
\langle\Psi(x')G^{(-)}(v,\Omega)\rangle&=\frac{i\left(\begin{array}{c}0\\\lambda^\dagger_{\dot a}\end{array}\right)}{8\pi^2(v-\bar n\cdot x'+i\epsilon)^2},\quad \langle\Psi(x')\bar F^{(-)}(v,\Omega)\rangle=\frac{i\left(\begin{array}{c}\lambda_a\\0\end{array}\right)}{8\pi^2(v-\bar n\cdot x'+i\epsilon)^2}.
 \end{align}\es We checked that they are consistent with \eqref{bulktoboudnr}. Similar to the relation between the Feynman propagator of scalar and fermion, 
 we can transform them to the scalar bulk-to-boundary propagator 
\begin{subequations}
\begin{align}
    \langle F(u,\Omega)\bar\Psi(x')\rangle &=i(0, \lambda_{\dot a}^\dagger)\partial_u D_S(u,\Omega;x'),\quad \langle \bar G(u,\Omega)\bar \Psi(x')\rangle=i(\lambda^a,0)\partial_u D_S(u,\Omega;x'),\\
\langle G(u,\Omega)\Psi(x')\rangle&=i\left(\begin{array}{c}0\\ \lambda^{\dagger\dot a}\end{array}\right)\partial_u D_S(u,\Omega;x'),\quad \langle \bar F(u,\Omega)\Psi(x')\rangle=i\left(\begin{array}{c}\lambda_a\\0\end{array}\right)\partial_u D_S(u,\Omega;x'),\\
\langle \bar\Psi(x')F^{(-)}(v,\Omega)\rangle&=-i(0,\lambda_{\dot a}^\dagger)\partial_v D^{(-)}_S(v,\Omega;x'),\quad \langle \bar \Psi(x')\bar G^{(-)}(v,\Omega)\rangle=-i(\lambda^a,0)\partial_v D^{(-)}_S(v,\Omega;x'),\\
\langle\Psi(x')G^{(-)}(v,\Omega)\rangle&=-i\left(\begin{array}{c}0\\\lambda^\dagger_{\dot a}\end{array}\right)\partial_v D^{(-)}_S(v,\Omega;x'),\quad \langle\Psi(x')\bar F^{(-)}(v,\Omega)\rangle=-i\left(\begin{array}{c}\lambda_a\\0\end{array}\right)\partial_v D^{(-)}_S(v,\Omega;x')
\end{align} where 
\bea 
D_S(u,\Omega;x')=-\frac{1}{8\pi^2(u+n\cdot x'-i\epsilon)},\quad D_S^{(-)}(v,\Omega;x')=\frac{1}{8\pi^2(v-\bar n\cdot x'+i\epsilon)}.
\eea 
\end{subequations}
 \item \textbf{Boundary-to-boundary propagators.} We can also derive the boundary-to-boundary correlators 
 \begin{subequations}\begin{align}
\langle F(u,\Omega)G^{(-)}(v',\Omega')\rangle&=0,\\
\langle F(u,\Omega)\bar F^{(-)}(v',\Omega')\rangle&=\frac{i}{4\pi(u-v'-i\epsilon)}\delta(\Omega-\Omega'),\\
\langle \bar G(u,\Omega)G^{(-)}(v',\Omega')\rangle&=\frac{i}{4\pi(u-v'-i\epsilon)}\delta(\Omega-\Omega'),\\
\langle \bar G(u,\Omega)\bar F^{(-)}(v',\Omega')\rangle&=0,\\
\langle G(u,\Omega)F^{(-)}(v',\Omega')\rangle&=0,\\
\langle \bar F(u,\Omega)F^{(-)}(v',\Omega')\rangle&=\frac{i}{4\pi(u-v'-i\epsilon)}\delta(\Omega-\Omega'),\\
\langle G(u,\Omega)\bar G^{(-)}(v',\Omega')\rangle&=\frac{i}{4\pi(u-v'-i\epsilon)}\delta(\Omega-\Omega'),\\
\langle \bar F(u,\Omega)\bar G^{(-)}(v',\Omega')\rangle&=0.
\end{align}\end{subequations} More compactly, they are equivalent to 
\begin{subequations}
\begin{align}
    \langle \psi(u,\Omega)\bar \psi^{(-)}(v',\Omega')\rangle&=\lim{}_-r' \langle \psi(u,\Omega)\bar\Psi(x')\rangle=-\frac{i\slashed n}{4\pi(u-v'-i\epsilon)}\delta(\Omega-\Omega'^{\text P}),\\ \langle \bar \psi(u,\Omega)\psi^{(-)}(v',\Omega')\rangle&=\lim{}_-r'\langle \bar \psi(u,\Omega)\Psi(x')\rangle=-\frac{i\slashed n}{4\pi(u-v'-i\epsilon)}\delta(\Omega-\Omega'^{\text P}).
\end{align} \end{subequations} Note that we have reduced the bulk-to-boundary propagators to the boundary-to-boundary propagator. The two-point correlator from bulk scalar theory can be found in \cite{Liu:2022mne}. Note that there is an additional  magnetic branch in the scalar theory, which is absent in the free fermion theory.
\item \textbf{Split representation.} The split representation is to express the Feynman propagator as
the product of bulk-to-boundary propagators. In AdS/CFT, this representation is derived in \cite{Costa:2014kfa}. In the context of Carrollian holography, similar split representation exists and has been derived for scalar theory \cite{Liu:2024nfc}. Notice that the Feynman propagator of Dirac spinor is related to one of the scalar theory via (we assume $x^0>x'^0$)
\bea 
\langle \Psi(x)\bar\Psi(x')\rangle=i\gamma^\mu\partial_\mu\langle\Phi(x)\Phi(x')\rangle=-2\int du d\Omega \slashed n \partial_u D_S^*(u,\Omega;x)\partial_u D_S(u,\Omega;x').
\eea The bulk-to-boundary propagators for the Dirac spinor \eqref{psibarPsi} and \eqref{psibarPsi2} are related to the scalar ones 
\bea 
\langle\psi(u,\Omega)\bar\Psi(x)\rangle=\langle \bar\psi(u,\Omega)\Psi(x)\rangle=-i\slashed n \partial_u D_s(u,\Omega;x).
\eea 
Therefore, the split representation for the Dirac propagator is 
\bea 
\langle\Psi(x)\bar\Psi(x')\rangle=\int du d\Omega \langle \Psi(x)\bar \psi(u,\Omega)\rangle \slashed{\bar n} \langle \psi(u,\Omega)\bar \Psi(x')\rangle
\eea where we have used the identity 
\be 
\slashed n \slashed{\bar n}\slashed n=\slashed n\slashed{\bar n}\slashed n+\slashed{\bar n}\slashed n\slashed n=-4 \slashed n.
\ee Note this is equivalent to the one-particle completeness relation 
\bea 
1=\frac{1}{2}\int du d\Omega |\bar\psi(u,\Omega)\rangle \slashed{\bar n}\langle \psi(u,\Omega)|.
\eea 
 \end{itemize}
 \subsection{Transformation laws}
 The transformation laws of the boundary fields can be induced from the bulk field. Poincar\'e symmetry $ISO(1,3)$ can be decomposed as the semi-direct product of the Lorentz transformations and the spacetime translation. For spacetime translation, the bulk and boundary coordinates transform as \eqref{translation} and the spinor transforms as 
 \be 
 \Psi'(x')=\Psi(x).
 \ee Therefore, we find 
 \be 
 F'(u',\Omega')=F(u,\Omega),\quad \overline G'(u',\Omega')=\overline G(u,\Omega),\quad \text{with}\quad u'=u-c\cdot n,\quad \Omega'=\Omega.
 \ee For Lorentz transformation of the bulk and boundary coordinates can be found in \eqref{ntranslation}. In particular, the null vector $n^\mu$ transforms as 
 \be 
 n'^\mu=\Gamma^{-1}\Lambda^\mu_{\ \nu}n^\nu.
 \ee Switching to the matrix form, we find 
 \bea 
 -\lambda'_a \lambda_{\dot a}^{'\dagger}=n'^\mu \sigma_{\mu a\dot a}=\Gamma^{-1}\Lambda^\mu_{ \nu}n^\nu \sigma_{\mu a\dot a}=\Gamma^{-1}\Lambda^\mu_{\ \nu}(-\frac{1}{2})\bar\sigma^{\nu\dot b b}\sigma_{\mu a\dot a}n_{b\dot b}.
 \eea At the first step, we still parameterize $n'_{a\dot a}$ as a pair of commuting spinors. At the last step, we have used the inverse relation 
 \be 
 n^\mu=-\frac{1}{2}\bar{\sigma}^{\mu \dot a a} n_{a\dot a}.
 \ee Note that the $\gamma^\mu$ matrices satisfy the identity 
\bea 
\Lambda_{-1/2}\gamma^\mu \Lambda_{1/2}=\Lambda^\mu_{\ \nu}\gamma^\nu
\quad\Leftrightarrow\quad 
\Lambda^L_{-1/2}\sigma^\mu \Lambda_{1/2}^R=\Lambda^\mu_{\ \nu}\sigma^\nu,\quad \Lambda_{-1/2}^R\bar\sigma^\mu \Lambda_{1/2}^L=\Lambda^\mu_{\ \nu}\bar\sigma^\nu\label{lorentzspinor}
\eea Then 
\bea 
\lambda_a'\lambda^{'\dagger}_{\dot a}=\Gamma^{-1}\left(\Lambda_{1/2}^L\right)_a^{\ b} \left(\Lambda_{-1/2}^R\right)^{\dot b}_{\ \dot a}\lambda_b\lambda_{\dot b}^\dagger.\label{lambdatransf}
\eea 
It follows that the transformation law of $\lambda_a$ and $\lambda_{\dot a}^\dagger$ would be
\bea 
\lambda_a'=\Gamma^{-1/2}\left(\Lambda_{1/2}^L\right)_a^{\ b}\lambda_b e^{i\text{phase}},\quad \lambda_{\dot a}^{'\dagger}=\Gamma^{-1/2}\lambda_{\dot b}^\dagger\left(\Lambda_{-1/2}^R\right)^{\dot b}_{\ \dot a}e^{-i\text{phase}}
\eea up to an unknown phase factor. To determine it, we use the definition of $\lambda_a$ in \eqref{deflambda} and find 
\bea 
\lambda_a'&=&i\sqrt{\frac{1}{1+|z'|^2}}(1,-z')=i\Gamma^{-1/2}\sqrt{\frac{1}{1+|z|^2}}|cz+d|(1,-\frac{az+b}{cz+d})=\Gamma^{-1/2}\sqrt{\frac{\bar c\bar z+\bar d}{cz+d}}\left(\begin{array}{cc}d&-c\\-b&a\end{array}\right)_a^{\ b}\lambda_b.\nn\\\eea 
The matrix $\left(\begin{array}{cc}\bar d&-\bar b\\-\bar c&\bar a\end{array}\right)$ is an element of the M\"{o}bius group and we may identify it as $\Lambda_{1/2}^L$
\bea 
\Lambda_{1/2}^L=\left(\begin{array}{cc}d&-c\\-b&a\end{array}\right)\Leftrightarrow\quad \Lambda_{-1/2}^L=\left(\begin{array}{cc}a&c\\b&d\end{array}\right).\label{lambdaL}
\eea Thus the phase factor is 
\be 
e^{i\text{phase}}=\sqrt{\frac{\bar c\bar z+\bar d}{cz+d}}.
\ee We will introduce  a $U(1)$ element $t$ \footnote{This notation follows \cite{Liu:2024llk} and should not be confused with Cartesian time.}
\be t=\frac{\bar c\bar z+\bar d}{cz+d}=\left(\frac{\partial z'}{\partial z}\right)^{1/2}\left(\frac{\partial \bar z'}{\partial\bar z}\right)^{-1/2}
\ee and the transformation law of $\lambda_a$ becomes 
\be 
\lambda'_a=\Gamma^{-1/2}t^{1/2}\left(\Lambda_{1/2}^L\right)_a^{\ b}\lambda_b.\label{translambda}
\ee Similarly, 
\bea 
\lambda'^\dagger_{\dot a}=\Gamma^{-1/2}t^{-1/2}\lambda_{\dot b}^\dagger \left(\Lambda_{-1/2}^R\right)^{\dot b}_{\ \dot a}\label{trawns}
\eea where 
\bea 
\Lambda_{-1/2}^R=\left(\begin{array}{cc}\bar d&-\bar b\\-\bar c&\bar a\end{array}\right)\Leftrightarrow\quad\Lambda_{1/2}^R=\left(\begin{array}{cc}\bar a&\bar b\\\bar c&\bar d\end{array}\right).\label{lamdaR}
\eea As a consistent check, we substitute \eqref{lambdaL} and \eqref{lamdaR} into \eqref{lorentzspinor} to  calculate the matrix $\Lambda^{\mu\nu}$\bea 
\Lambda^{\mu\nu}=-\frac{1}{2}\text{tr}\left(\Lambda_{-1/2}^L\sigma^\mu \Lambda_{1/2}^R\bar\sigma^\nu\right).
\eea More precisely, 
\bea 
\footnotesize
\Lambda^{\mu\nu}=-
\frac{1}{2}\left(
\begin{array}{cccc}
 a \bar{a}+b \bar{b}+c \bar{c}+d \bar{d} & -b \bar{a}-a \bar{b}-d \bar{c}-c \bar{d} & i \left(b \bar{a}-a \bar{b}+d \bar{c}-c \bar{d}\right) & -a \bar{a}+b \bar{b}-c \bar{c}+d \bar{d} \\
 c \bar{a}+a \bar{c}+d \bar{b}+b \bar{d} & -d \bar{a}-a \bar{d}-c \bar{b}-b \bar{c} & i \left(d \bar{a}-a \bar{d}-c \bar{b}+b \bar{c}\right) & -c \bar{a}-a \bar{c}+d \bar{b}+b \bar{d} \\
 i \left(c \bar{a}-a \bar{c}+d \bar{b}-b \bar{d}\right) & i \left(-d \bar{a}+a \bar{d}-c \bar{b}+b \bar{c}\right) & -d \bar{a}-a \bar{d}+c \bar{b}+b \bar{c} & i \left(-c \bar{a}+a \bar{c}+d \bar{b}-b \bar{d}\right) \\
 a \bar{a}+b \bar{b}-c \bar{c}-d \bar{d} & -b \bar{a}-a \bar{b}+d \bar{c}+c \bar{d} & i \left(b \bar{a}-a \bar{b}-d \bar{c}+c \bar{d}\right) & -a \bar{a}+b \bar{b}+c \bar{c}-d \bar{d} \\
\end{array}
\right)\nn\\
\eea and 
\bea 
\footnotesize
\Lambda^\mu_{\ \nu}= -\frac{1}{2}\left(
\begin{array}{cccc}
 -a \bar{a}-b \bar{b}-c \bar{c}-d \bar{d} & -b \bar{a}-a \bar{b}-d \bar{c}-c \bar{d} & i \left(b \bar{a}-a \bar{b}+d \bar{c}-c \bar{d}\right) & -a \bar{a}+b \bar{b}-c \bar{c}+d \bar{d} \\
 -c \bar{a}-a \bar{c}-d \bar{b}-b \bar{d} & -d \bar{a}-a \bar{d}-c \bar{b}-b \bar{c} & i \left(d \bar{a}-a \bar{d}-c \bar{b}+b \bar{c}\right) & -c \bar{a}-a \bar{c}+d \bar{b}+b \bar{d} \\
 i \left(-c \bar{a}+a \bar{c}-d \bar{b}+b \bar{d}\right) & i \left(-d \bar{a}+a \bar{d}-c \bar{b}+b \bar{c}\right) & -d \bar{a}-a \bar{d}+c \bar{b}+b \bar{c} & i \left(-c \bar{a}+a \bar{c}+d \bar{b}-b \bar{d}\right) \\
 -a \bar{a}-b \bar{b}+c \bar{c}+d \bar{d} & -b \bar{a}-a \bar{b}+d \bar{c}+c \bar{d} & i \left(b \bar{a}-a \bar{b}-d \bar{c}+c \bar{d}\right) & -a \bar{a}+b \bar{b}+c \bar{c}-d \bar{d} \\
\end{array}
\right).\nn\\
\eea This matches with the result in \cite{Penrose:1985bww}.
Using the fall-off condition of the spinor field near $\mathcal I^+$, we find 
\bea 
\frac{\psi'(u',\Omega')}{r'}+\cdots=\Lambda_{1/2} \frac{\psi(u,\Omega)}{r}+\cdots.
\eea Since $r'=\Gamma r$ under Lorentz transformations, the leading coefficient should satisfy the relation 
\be 
\psi'(u',\Omega')=\Gamma \Lambda_{1/2}\psi(u,\Omega).
\ee Combining with the transformation law of the commuting spinors \eqref{translambda} and \eqref{trawns}, we find \footnote{The infinitesimal transformation 
\bea 
\delta_Y F&=&F'(u,z,\bar z)-F(u,z,\bar z)=-\frac{1}{2} \nabla_AY^A u\dot F-\frac{3}{4}\nabla_AY^A F-Y^A\nabla_A F-\frac{1}{4}(\partial_zY^z-\partial_{\bar z}Y^{\bar z})F
\eea is slightly different  from eqn. (2.76a) of \cite{Guo:2024qzv}. This is because the choices of the commuting spinors are differed by a local phase. This local rotation is allowed due to the little group scaling. One can find similar ambiguity in the vector and gravitational theory where it is called superduality \cite{Liu:2024llk,Liu:2023qtr,Liu:2023gwa}. Its relation to helicity and topological invariants is discussed in \cite{Liu:2024rvz,Long:2025fbb} and applications to real systems can be found in \cite{Long:2024yvj,Heng:2025kmr}. }
\be 
F'(u',\Omega')=\Gamma^{3/2}t^{-1/2}F(u,\Omega),\quad \bar G'(u',\Omega')=\Gamma^{3/2}t^{1/2}\bar G(u,\Omega).\label{fertrans}
\ee The factor $\Gamma^{3/2}$  arises from the transformation law of the commuting spinors ($\Gamma^{1/2}$) and the fall-off behavior ($\Gamma$). In \cite{Liu:2024llk}, a bulk field $\Phi_s(x)$ with spin s  is related to the boundary field $O_s(u,\Omega)$ via the fall-off condition 
\be 
\Phi_s(x)=\frac{O_s(u,\Omega)}{r}+\cdots
\ee and $O_s$ is identified as a fundamental field at the boundary. For $s=0$, $O_0$ is already a primary field with spin 0. For $s=1,2,\cdots$, one can extract two transverse modes $O_{\pm s}$ from $O_s$ that correspond to the two propagating degrees of freedom. The field $O_{s}$ transforms as follows under Lorentz rotations 
\bea 
O_s'(u',\Omega')=\Gamma t^s O_s(u,\Omega),\quad s=0,\pm 1,\pm 2,\cdots.\label{bostrans}
\eea We can unify the transformation laws of the  fermionic operators in \eqref{fertrans} and the bosonic operators in \eqref{bostrans} as follows. A primary field $O_{\bar\Delta,s}(u,\Omega)$ on $\mathcal I^+$ with conformal weight $\bar\Delta$ and helicity $s$ is defined as 
\bea 
O'_{\bar\Delta,s}(u',\Omega')=\Gamma^{\bar\Delta}t^sO_{\bar\Delta,s}(u,\Omega).\label{primary}
\eea Note that the conformal weight $\bar\Delta$ is not always equal to the fall-off index in the bulk. Therefore, we use the notation $\bar\Delta$ to distinguish it with $\Delta$. Introducing two alternative quantum numbers $(h,\bar h)$ via the relation
\be 
\bar\Delta=-h-\bar h,\quad s=h-\bar h.
\ee Then the transformation law can be written in stereographic coordinates
\bea 
O'_{h,\bar h}(u',z',\bar z')=\left(\frac{1+|z'|^2}{1+|z|^2}\right)^{-h-\bar h}\left(\frac{\partial z'}{\partial z}\right)^{h}\left(\frac{\partial\bar z'}{\partial \bar z}\right)^{\bar h}O_{h,\bar h}(u,z,\bar z)
\eea where 
\bea 
u'=\Gamma^{-1}u,\quad z'=\frac{az+b}{cz+d}.
\eea The result can be compared with the one in \cite{Banerjee:2018gce}. The factor $\left(\frac{1+|z'|^2}{1+|z|^2}\right)^{-h-\bar h}$ comes from the fact that the asymptotic null manifold is topologically $\mathbb R\times S^2$. Once the  Carrollian manifold  is topologically $\mathbb R\times\mathbb C$, this annoying factor disappears.  
\paragraph{Remarks.} 
\begin{itemize}
    \item Our result establishes a direct connection between the bulk fields and the extrapolating fields on the boundary. For low spins, the quantum numbers are summarized in table \ref{qun}. For the fields with non-vanishing spin, there should be two primary operators that correspond to the two physical polarization states in the bulk.
     \begin{table}
\centering 
\renewcommand{\arraystretch}{1.5}
\begin{tabular}{|c|c|c|c|c|}\hline
Bulk field&Real scalar field&Dirac spinor&Maxwell field&Gravitational field\\\hline
Boundary field $(h,\bar h)$&$(-\frac{1}{2},-\frac{1}{2})$&$(-\frac{1}{2},-1),(-1,-\frac{1}{2})$&$(0,-1),(-1,0)$&$(\frac{1}{2},-\frac{3}{2}),(-\frac{3}{2},\frac{1}{2})$\\\hline
\end{tabular}
\caption{\centering{Quantum numbers for lower spins.}}
\label{qun}
\end{table}
\item In all the examples we have considered, the conformal weight $\bar\Delta$ is always no smaller than the fall-off index
\be 
\bar\Delta\ge\Delta.\label{ch}
\ee Since we do not expect the bulk field to be divergent near $\mathcal I^+$, it is reasonable to impose the condition $\Delta\ge 0$. Therefore, once the inequality \eqref{ch} holds, we find a constraint on the summation of quantum numbers $h$ and $\bar h$
\be 
h+\bar h\le 0.
\ee Furthermore, as the helicity $s$ is always an integer or a half integer, we may require 
\bea 
h-\bar h\in \left\{\begin{array}{cc}\mathbb Z&\ \text{bosonic field}\\ \mathbb Z+\frac{1}{2}&\ \text{fermionic field}.\end{array}\right.
\eea 
\item For a general Carrollian correlator, the spacetime invariance leads to the identity  
\bea 
\langle \prod_{j=1}^m O_{\bar\Delta_j,s_j}(u'_j,\Omega_j)\rangle=\langle \prod_{j=1}^m O_{\bar\Delta_j,s_j}(u_j,\Omega_j)\rangle.
\eea Similarly, the Ward identity for the Lorentz symmetry is 
\bea 
\langle \prod_{j=1}^m O_{\bar\Delta_j,s_j}(u'_j,\Omega'_j)\rangle=\left(\prod_{j=1}^m \Gamma_j^{\bar\Delta_j}t_j^{s_j}\right)\langle\prod_{j=1}^m O_{\bar\Delta_j,s_j}(u_j,\Omega_j)\rangle.\label{ls}
\eea 
\end{itemize}
 \subsection{Four-point Carrollian amplitudes with fermionic operators}
Since the two-point and three-point Carrollian amplitudes are fixed to several known structures,  we will use the formula \eqref{caf} to derive several four-point Carrollian amplitudes with fermionic operators in this subsection. We will use the spinor helicity formalism that is reviewed  in \cite{2015sagt.book.....E}. A momentum $p^\mu$ is mapped to a $2\times 2$ matrix that can be decomposed into angle and square spinors
\be 
p_{a\dot a}=p^\mu \sigma_{\mu a\dot a}=-|p]_a\langle p|_{\dot a}.
\ee Comparing with \eqref{padota}, we find the one-to-one correspondence up to an overall little group scaling
\be 
|p]_a=\sqrt{\omega}\lambda_a,\quad \langle p|_{\dot a}=\sqrt{\omega}\lambda_{\dot a}^\dagger.
\ee  The angle/square spinor product are defined using the invariant symbol $\epsilon_{\dot a\dot b}$ and $\epsilon_{ab}$
\bs\begin{align}
[ 12]&=[p_1p_2]=\sqrt{\omega_1\omega_2}\lambda_1^a \lambda_{2a}=\frac{2z_{12}\sqrt{\omega_1\omega_2}}{\sqrt{(1+|z_1|^2)(1+|z_2|^2)}},\\
\langle 12\rangle&=\langle p_1p_2\rangle=\sqrt{\omega_1\omega_2}\lambda_{1\dot a}\lambda_{2}^{\dot a}=-\frac{2\bar z_{12}\sqrt{\omega_1\omega_2}}{\sqrt{(1+|z_1|^2)(1+|z_2|^2)}}
\end{align}\es where $z_{ij}=z_i-z_j,\ \bar z_{ij}=\bar z_i-\bar z_j$.
 
\paragraph{Yukawa theory}
Consider a Dirac fermion $\Psi$ interacting with a real scalar $\Phi$ via a Yukawa coupling. The 
4-fermion tree level scattering amplitude is $\mathcal A_4(\bar f^{h_1}f^{h_2}\bar f^{h_3}f^{h_4})$ where $f$ denotes an outgoing fermion and $\bar f$ an outgoing anti-fermion. 
The superscripts indicate the helicity.  Suppose we take particles
1 and 2 to have negative helicity and 3 and 4 positive. Then the  4-fermion tree amplitude is
\bea
\mathcal A_{-,-,+,+}[1,2,3,4]\equiv \mathcal A_4(\bar f^{-}f^{-}\bar f^{+}f^{+})=ig^2\frac{[34]}{[12]}=ig^2\frac{\langle12\rangle}{\langle34\rangle}
\eea where $g$ is the Yukawa coupling constant. The four-point Carrollian amplitude \eqref{caf} is 
\bea 
&&\mathcal C_{-,-,+,+}[1,2,3,4]\nn\\&=&\langle \prod_{j=4}^4 \uppsi_{s_j}(u_j,\Omega_j,\sigma_j,a_j)\rangle\nn\\&=&ig^2\left(\frac{1}{8\pi^2}\right)^4 \int_0^\infty d\omega_1 d\omega_2 d\omega_3 d\omega_4 \sqrt{\omega_1\omega_2\omega_3\omega_4}e^{-i\sum_{j=1}^4\sigma_j\omega_ju_j}\frac{\langle 12\rangle}{\langle 34\rangle}(2\pi)^4\delta^{(4)}(\sum_{j=1}^4 p_j).
\eea We have inserted a Dirac delta function into the integrand due to the conservation of four-momentum. The momentum $p_j=\sigma_j \omega_j n_j$. The Lorentz tranformation law for the angle spinor bracket is 
\be 
\langle 12\rangle\to \langle 1'2'\rangle=\left(t_1^{*}t_2^*\right)^{1/2}\langle 12\rangle.
\ee As a consequence, the scattering amplitude transforms as follows
\bea 
\mathcal A_{-,-,+,+}[1',2',3',4']=\left(t_1^*t_2^*t_3 t_4\right)^{1/2}\mathcal A_{-,-,+,+}[1,2,3,4].
\eea Therefore, the Lorentz transformation law of the four-point Carrollian amplitude is 
\be 
C'_{-,-,+,+}[1',2',3',4']=\left(\Gamma_1\Gamma_2\Gamma_3\Gamma_4\right)^{3/2}\left(t_1^*t_2^*t_3 t_4\right)^{1/2} C_{-,-,+,+}[1,2,3,4].\label{Cmmpp}
\ee The  Lorentz transformation law of $u$ is $u'=\Gamma^{-1}u$. Therefore, one may introduce $\omega'=\Gamma \omega$ to preserve the phase in the integrand. Thus, we find a factor $\Gamma^{3/2}$ for each boundary field.  The transformation law \eqref{Cmmpp} agrees with \eqref{ls}.

For the four-point amplitude, we can fix $z_1=0,\ z_2=1,\ z_3=\infty,\ z_4=z$ by Lorentz transformations. In this special case, we have
\begin{align}
    \langle 12\rangle=\sqrt{2\omega_1\omega_2},\quad
    \langle 34\rangle=-\sqrt{\frac{4\omega_3\omega_4}{1+|z|^2}}.
\end{align} In this paper, we will always choose $\sigma_1=\sigma_2=-,\ \sigma_3=\sigma_4=+$. The Dirac delta function can be reduced to 
\be 
\delta^{(4)}(\sum_{j=1}^4p_j)=\frac{1+z^2}{2\omega_4}\delta(\omega_1-\frac{1-z}{1+z^2}\omega_4)\delta(\omega_2-\frac{2z}{1+z^2}\omega_4)\delta(\omega_3-\frac{z(1-z)}{1+z^2}\omega_4)\delta(\bar z-z)
\ee and then the Carrollian amplitude becomes 
\bea 
{C}_{-,-,+,+}[1,2,3,4]&=&-\Theta(z)\Theta(1-z)\delta(\bar z-z)\frac{ig^2}{(4\pi)^4\sqrt{2}}\frac{z(1-z)}{\sqrt{1+z^2}}\int_0^\infty d\omega_4 \omega_4 e^{-i\omega_4\chi}\nn\\&=&\frac{ig^2z(1-z)}{\sqrt{2(1+z^2)}}\Theta(z)\Theta(1-z)\delta(\bar z-z) \chi^{-2},
\eea
where 
\bea 
\chi=u_4-\frac{1-z}{1+z^2}u_1-\frac{2z}{1+z^2}u_2+\frac{z(1-z)}{1+z^2}u_3
\eea is translation invariant. Interestingly, the integral for $\omega_4$ in the four-fermion Carrollian amplitude is in the form of 
\be 
\int_0^\infty d\omega_4 \omega_4 e^{-i\omega_4\chi}\label{omega4}
\ee which is finite while in the scalar case this is 
\be 
\int_0^\infty d\omega_4 \omega_4^{-1}e^{-i\omega_4\chi}.
\ee The latter is divergent near $\omega_4=0$ and one should regularize it. The expression \eqref{omega4} can be traced back to the factors $\sqrt{\omega}$ that avoids divergences in \eqref{caf}. The result can be transformed back to the general reference frame using Lorentz transformation law. This has been done explicitly in \cite{Liu:2024nfc,Liu:2024llk}, we will not repeat it here.  

Next we turn to study the tree scattering among  two scalars and two fermions in Yukawa theory. The scattering amplitude in bulk is denoted as $\mathcal{A}_4(\Phi \bar f^{h_2}  f^{h_3} \Phi)$. The amplitude vanishes once the fermions have the same helicity. Thus the non-vanishing scattering amplitude is 
\bs\begin{align}
\mathcal A_{0,-,+,0}[1,2,3,4]&=\mathcal{A}_4(\phi \bar f^{-}  f^{+} \phi)=-ig^2(\frac{\langle13 \rangle }{\langle12 \rangle}+\frac{\langle34 \rangle }{\langle 24\rangle}),\\
\mathcal A_{0,+,-,0}[1,2,3,4]&=\mathcal{A}_4(\phi \bar f^{+}  f^{-} \phi)=-ig^2(\frac{[13] }{[12]}+\frac{[34] }{[24]})
\end{align}\es 
and we will only consider the first one. The four-point Carrollian amplitude is slightly different from \eqref{caf} since there are two real scalars in the process
\bea 
&&C_{0,-,+,0}[1,2,3,4]\nn\\&=&\left(\frac{1}{8\pi^2 i}\right)^4 \int_0^\infty d\omega_1 \int_0^\infty d\omega_2 \sqrt{\omega_2} \int_0^\infty d\omega_3\sqrt{\omega}_3 \int_0^\infty d\omega_4 e^{-i\sum_{j=1}^4\sigma_j\omega_j u_j}\nn\\
&\times&\mathcal A_{0,-,+,0}[1,2,3,4](2\pi)^4\delta^{(4)}(\sum_{j=4}^4 p_j).\nn\\
\eea In the integral transform, there is a factor of $\sqrt{\omega}$ for each fermion and a factor 1 for each scalar. 
We still fix $z_1=0,\ z_2=1,\ z_3=\infty,\ z_4=z$ and then the Carrollian amplitude is 
\bea 
C_{0,-,+,0}[1,2,3,4]=-\frac{g^2}{(4\pi)^4\sqrt{2}}z(2-z)\Theta(z)\Theta(1-z)\delta(\bar z-z)\chi^{-1}.
\eea Again, the integral is convergent due to the insersion of $\sqrt{\omega_2}$ and $\sqrt{\omega_3}$.

\paragraph{Massless QED}
Now we consider QED Compton scattering with two fermions and two photons. The scattering amplitude is denoted as $\mathcal{A}_4(\bar f^{h_1} f^{h_2} \gamma^{h_3} \gamma^{h_4})$ where $\gamma^h$ represents a photon with helicity $h$.  If two fermions have the same helicity,the amplitude will vanish. The condition is the same to the photons. So a non-vanishing amplitude is
\bea
\mathcal A_{+,-,+,-}[1,2,3,4]\equiv \mathcal{A}_4(\bar f^+ f^- \gamma^+ \gamma^-)=ie^2 \frac{\langle24 \rangle^2}{\langle13 \rangle \langle23 \rangle},
\eea and the Carrollian amplitude is 
\bea 
&&C_{+,-,+,-}[1,2,3,4]\nn\\&=&\left(\frac{1}{8\pi^2 i}\right)^4 \int_0^\infty d\omega_1\sqrt{\omega_1} \int_0^\infty d\omega_2 \sqrt{\omega_2}\int_0^\infty d\omega_3 \int_0^\infty d\omega_4 e^{-i\sum_{j=1}^4\sigma_j\omega_j u_j}\nn\\
&\times& \mathcal A_{+,-,+,-}[1,2,3,4](2\pi)^4\delta^{(4)}(\sum_{j=4}^4 p_j).\nn\\
\eea Similar to the scattering with two scalars and two fermions, we should insert a factor $\sqrt{\omega}$ for each fermion and insert a 1 for each photon. The following computation is straightforward
\bea 
C_{+,-,+,-}[1,2,3,4]=-\frac{e^2}{(4\pi)^4\sqrt{2}}(1-z)\Theta(z)\Theta(1-z)\delta(\bar z-z)\chi^{-1}.
\eea 

\section{Extrapolating limits}\label{el} 
\begin{figure}
\begin{center}
\begin{tikzpicture}[scale=1.2,>=stealth,
    boundary/.style={thick},
    curve/.style={thick},
    dot/.style={fill,circle,inner sep=1.5pt}
]

\coordinate (iplusL) at (0,3);
\coordinate (i0L)   at (3,0);
\coordinate (iminusL) at (0,-3);

\draw[boundary] (iplusL) -- (i0L) node[midway,right=7pt] {$$};
\draw[boundary] (i0L) -- (iminusL) node[midway,right=9pt] {$$};
\draw[boundary] (iminusL) -- (iplusL);

\node[above] at (iplusL) {$i^+$};
\node[right] at (i0L)   {$i^0$};
\node[below] at (iminusL) {$i^-$};

\coordinate (SigmaUL) at (1.2,1.8);
\coordinate (SigmaDL) at (1.2,-1.8);
\node[dot] at (SigmaUL) {};
\node[dot] at (SigmaDL) {};
\node[right] at (SigmaUL) {$(u,\Omega)$};
\node[right] at (SigmaDL) {$(v',\Omega')$};

\draw[curve] (SigmaDL) to[out=70,in=-70] (SigmaUL);
\draw[curve] (SigmaDL) to[out=110,in=-110] (SigmaUL);

\coordinate (iplusR) at (6,3);
\coordinate (i0R)   at (9,0);
\coordinate (iminusR) at (6,-3);

\draw[boundary] (iplusR) -- (i0R) node[midway,right=7pt] {$$};
\draw[boundary] (i0R) -- (iminusR) node[midway,right=9pt] {$$};
\draw[boundary] (iminusR) -- (iplusR);

\node[above] at (iplusR) {$i^+$};
\node[right] at (i0R)   {$i^0$};
\node[below] at (iminusR) {$i^-$};

\coordinate (SigmaUR) at (7.2,1.8);
\coordinate (SigmaDR) at (7.2,-1.8);
\node[dot] at (SigmaUR) {};
\node[dot] at (SigmaDR) {};
\node[right] at (SigmaUR) {$(u,\Omega)$};
\node[right] at (SigmaDR) {$(v',\Omega')$};

\coordinate (X) at (7.2,0);
\node[dot] at (X) {}; 
\node[anchor=west, xshift=2pt] at (X) {$x$}; 

\draw[curve] (X) to[out=50,in=-70] (SigmaUR);  
\draw[curve] (X) to[out=130,in=-110] (SigmaUR); 
\draw[curve] (X) to[out=-50,in=70] (SigmaDR);  
\draw[curve] (X) to[out=-130,in=110] (SigmaDR); 

\end{tikzpicture}
\end{center}\caption{The leading the subleading contributions to the two-point correlation function of the composite operator $\Sigma^2$. }\label{phi2}
\end{figure}
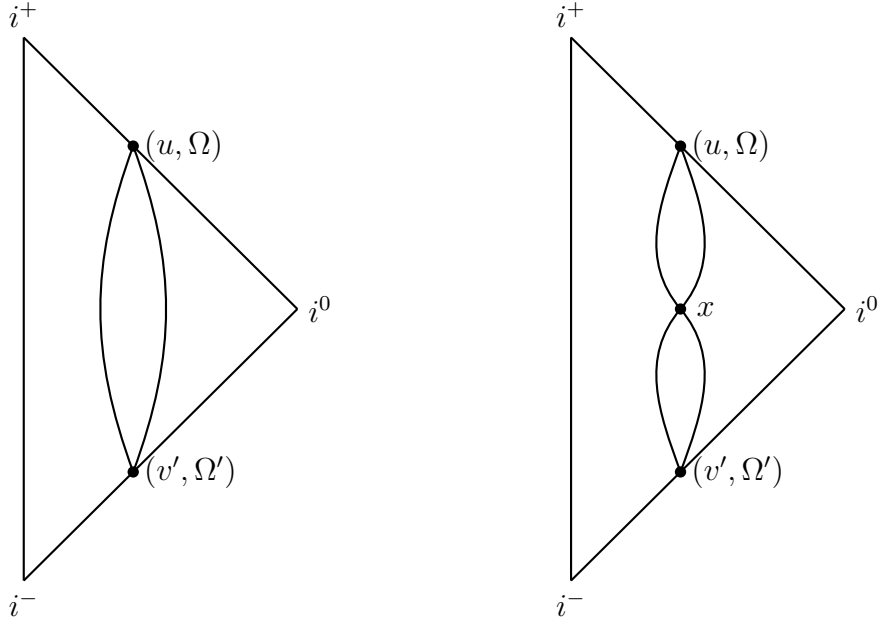

As the bulk-to-boundary correlator is fixed by symmetry, one can extrapolate the remaining bulk point to the boundary and define the boundary-to-boundary correlator. This method has been used for extracting finite temperature Carrollian correlators in \cite{Long:2025bfi}.  For the scalar branch, we expect \footnote{$D_s$ denotes a scalar branch bulk-to-boundary correlator in any theory.}
     \be 
     B_s(u,\Omega;v',\Omega')=\lim{}_- r'^{\Delta} D_s(u,\Omega;x')=C_s\lim{}_-\frac{ r'^\Delta }{(u+n\cdot x'-i\epsilon)^{\Delta}}.
     \ee Since \be 
u+n\cdot x'=u-v'+r' n\cdot \bar n',
\ee there is always a magnetic branch when $\Omega\not=\Omega'^{\text P}$. We may write $B_s$ as a summation of an electric branch and a magnetic branch 
\be 
B_s(u,\Omega;v',\Omega')=\alpha(u-v')\delta(\Omega-\Omega'^{\text P})+\frac{C_s}{(1+\cos\gamma(\Omega,\Omega')-i\eta)^{\Delta}}
\ee where $\gamma(\Omega,\Omega')$ is the angle between $\bm n$ and $\bm n'$. The small parameter $\eta$ is related to $\epsilon$ via $\eta=\frac{\epsilon}{r'}$ and $\alpha(u-v')$ characterizes the electric branch which is determined by integrating both sides on the unit sphere. Interestingly, we find 
\bea 
\alpha(u-v')=\left\{\begin{array}{cc} 0&\ \Delta<1,\\
-2\pi C_s\ln i(u-v'-i\epsilon)/\epsilon&\ \Delta=1,\\
\text{divergent}&\Delta>1.\end{array}\right.\label{elecsc}
\eea We conclude that there is no electric branch for $\Delta<1$ and only the magnetic branch is important 
\be 
B_s(u,\Omega;v',\Omega')=\frac{C_s}{(1+\cos\gamma(\Omega,\Omega')-i\eta)^{\Delta}},\quad 0<\Delta<1.
\ee The  electric and the magnetic branch coexist for $\Delta=1$. Furthermore, their relative normalization coefficient are not independent 
\be 
B_s(u,\Omega;v',\Omega')=-2\pi C_s\ln i(u-v'-i\epsilon)/\epsilon{\delta(\Omega-\Omega'^{\text P})}+\frac{C_s}{1+\cos\gamma(\Omega,\Omega')-i\eta},\quad \Delta=1.
\ee Finally, the electric branch would be divergent for $\Delta>1$.  To be more precise, the left-hand side contributes
\bea 
&&\frac{2\pi C_s}{1-\Delta}r'^{\Delta-1}\left[\frac{1}{(u-v'+2r'-i\epsilon)^{\Delta-1}}-\frac{1}{(u-v'-i\epsilon)^{\Delta-1}}\right]\nn\\&=&-\frac{2\pi C_s}{1-\Delta}\left(\frac{r'}{u-v'-i\epsilon}\right)^{\Delta-1}+\frac{2\pi C_s}{1-\Delta}2^{1-\Delta}.
\eea The magnetic branch on the right-hand side  contributes 
\bea 
&&\frac{2\pi C_s}{1-\Delta}\left[\frac{1}{(2-i\eta)^{\Delta-1}}-\frac{1}{(-i\eta)^{\Delta-1}}\right]\nn\\&=& -\frac{2\pi C_s}{1-\Delta}\left(\frac{r'}{-i\epsilon}\right)^{\Delta-1}+\frac{2\pi C_s}{1-\Delta}2^{1-\Delta}.
\eea The constants on both sides cancel with each other. However, the divergent part does not cancel and thus $\alpha(u-v')$ suffers infrared divergence 
\be 
\alpha(u-v')\sim r'^{\Delta-1}.
\ee 
However, this does not mean that the electric branch is unphysical for $\Delta>1$. As an illustration, we consider the correlator of the composite operator $\Sigma^2$ in $\Phi^4$ theory. The leading and subleading diagrams are shown in figure \ref{phi2}. For the first diagram, one can use the boundary-to-boundary propagator of $\Sigma$ to obtain  
\be 
\langle\Sigma^2(u,\Omega)\Sigma^{2}(v',\Omega')\rangle\Big|_{\text{leading}}\sim \langle \Sigma(u,\Omega)\Sigma(v',\Omega')\rangle^2\sim \delta(\Omega-\Omega'^{\text P})\delta(0).
\ee The $\delta(0)$ is the Dirac delta function on $S^2$ whose argument is equal to zero. Obviously, this is divergent. For the second diagram, one can find 
\bea 
\langle\Sigma^2(u,\Omega)\Sigma^{2}(v',\Omega')\rangle\Big|_{\text{subleading}}\sim -i\lambda \int d^4 x \frac{1}{(u+n\cdot x-i\epsilon)^2(v'-\bar n'\cdot x+i\epsilon)^2}.
\eea 
Using the integral representation in frequency space, one can find that the correlator is non-vanishing only for $\Omega=\Omega'^{\text P}$. Moreover, the integral becomes
\bea 
\int_0^\infty d\omega \int_0^\infty d\omega' \omega \omega' e^{-i\omega u+i\omega' v'}\delta(\omega-\omega')\delta^{(3)}(\bm p+\bm p')\sim \frac{\widetilde\delta(0)}{u-v'-i\epsilon}\delta(\Omega-\Omega'^{\text P}).
\eea Here $\widetilde\delta(0)$ is the Dirac delta function 
\be 
\widetilde\delta(0)=\delta(\omega-\omega')\Big|_{\omega'=\omega}.
\ee Note that this is also divergent. One may conclude that the infrared divergence is inevitable for $\Delta>1$. The infrared divergence always appears in a massless theory and it does not mean the theory is problematic. Instead, usually one should redefine the operators to extract the real observables in the theory  \cite{1995iqft.book.....P}. We will leave the regularization of the infrared divergence in Carrollian theory as a topic for future study.

Now we turn to the fermionic branch and define the boundary-to-boundary correlator 
\be 
B_f(u,\Omega;v',\Omega')=\lim{}_- r'^\Delta D_f(u,\Omega;x')=C_f\lim{}_- \frac{r'^\Delta}{(u+n\cdot x'-i\epsilon)^{\Delta+1}}.
\ee When $\Omega\not=\Omega'^{\text P}$, 
\be 
B_f(u,\Omega;v',\Omega')=\alpha(u-v')\delta(\Omega-\Omega'^{\text P}).
\ee It is straightforward to obtain 
\be 
\alpha(u-v')=\left\{\begin{array}{cc}0&0 <\Delta<1,\\ 
2\pi C_f \slashed n (u-v'-i\epsilon)^{-1}&\Delta=1,\\
\text{divergent}&\Delta>1.\end{array}\right.
\ee When $0< \Delta<1$, the integral on $S^2$ leads to 
\bea 
-\frac{2\pi}{\Delta}r'^{\Delta-1}\left[\frac{1}{(u-v'+2r'-i\epsilon)^{\Delta}}-\frac{1}{(u-v'-i\epsilon)^{\Delta}}\right].\label{limit}
\eea The limit $r'\to \infty$ leads to a vanishing result. Therefore, we conclude that the boundary-to-boundary correlator is zero for $0< \Delta<1$
\be 
B_f(u,\Omega;v',\Omega')=0,\quad 0< \Delta<1.
\ee When $\Delta=1$, the boundary-to-boundary correlator is non-vanishing
\bea 
B_f(u,\Omega;v',\Omega')=\frac{2\pi C_f\slashed n}{u-v'-i\epsilon}\delta(\Omega-\Omega'^{\text P}).
\eea When $\Delta>1$, the expression \eqref{limit} is divergent and thus 
\be 
\alpha(u-v')\sim r'^{\Delta-1}.\label{alphadiv}
\ee 
In summary, the boundary-to-boundary correlator extrapolated from the bulk-to-boundary correlator has the general properties: 
\begin{itemize}
    \item There is no magnetic branch in fermionic branch for any $\Delta$. The magnetic branch only appears in the scalar branch. 
    \item When $0< \Delta<1$, there is only magnetic branch in the scalar branch 
    \be 
    B_s(u,\Omega;v',\Omega')=\frac{C_s}{(1+\cos\gamma(\Omega,\Omega')-i\eta)^{\Delta}}.
    \ee On the other hand, the boundary-to-boundary correlator is always zero in the fermionic branch.
    \item When $\Delta=1$, the magnetic branch and the electric branch coexist in the scalar boundary-to-boundary correlator. However, there is only an electric branch in the fermionic branch.
    \item When $\Delta>1$, the electric branch is always divergent both for  the scalar and fermionic branches. There may be a way to regularize it. 
\end{itemize}
\paragraph{Comparison with the other extrapolating method in the literature:}
In the above discussion, we have reduced the bulk-to-bulk correlator to  the bulk-to-boundary correlator, and then reduced the bulk-to-boundary correlator to the boundary-to-boundary correlator, namely
\bea
B_s(u,\Omega;v',\Omega')=\lim{}_- r'^{\Delta} D_s(u,\Omega;x')=\lim{}_- r'^{\Delta}\lim{}_+r^{\Delta}G(x;x')\label{ours}
\eea
in which we first take $r \to \infty$ and keep $u$ finite,  then  take $r' \to \infty$ and keep $v$ finite  to get the final boundary-to-boundary propagator.
While there is another extrapolation method
that can directly obtain the boundary-to-boundary correlator in one step from bulk-to-bulk correlator, e.g.\cite{Nguyen:2023miw}
\bea
B_s(u,\Omega;v',\Omega)=\lim_{r \to \infty}r^{2\Delta}G(x;x')\label{RRb2b}
\eea
in which it takes $r=-r' \to \infty$ in the meantime. The boundary-to-boundary correlator can be calculated when $G(x;x')$ is known explicitly. For a conformal field theory, one can verify that formula \eqref{RRb2b} yields a result compatible with \eqref{ours}. More precisely, the electric branch is absent for $0 < \Delta < 1$, while the electric and magnetic branches coexist at $\Delta = 1$. For $\Delta > 1$, however, the electric branch diverges as $r^{2\Delta-2}$, in contrast to \eqref{alphadiv}. Consequently, results are reliable only for $0 < \Delta \le 1$ while divergences for $\Delta > 1$ require regularization that yields scheme-independent results. 


However, a crucial distinction remains between formulae \eqref{RRb2b} and \eqref{ours}. Formula \eqref{RRb2b} cannot by itself yield the boundary-to-boundary correlator, as it requires additional input. Namely, the bulk-to-bulk correlator is not fixed by Poincar\'e symmetry. In contrast, formula \eqref{ours} incorporates the structure of the bulk-to-boundary correlator, which is symmetry determined. Consequently, it  fixes the boundary-to-boundary correlator for any massless theory, independently of any details regarding the bulk-to-bulk propagator.

\section{Discussions}
In this work, we have used the Ward identities to constrain the two-point bulk-to-boundary correlators in the theories with Poincar\'e symmetry. We find that the correlator is fixed up to a normalization constant for the scalar operator
\be 
D_s(u,\Omega;x')=\frac{C_s}{(u+n\cdot x'-i\epsilon)^\Delta}.\label{sbra}
\ee A bulk-to-boundary correlator of this form is called the correlator of the scalar branch. For fermionic fields in parity invariant theories, we find two possible branches. In addition to a scalar-like branch, there exists a distinct fermionic branch
\be 
D_f(u,\Omega;x')=\frac{C_f\slashed n}{(u+n\cdot x'-i\epsilon)^{\Delta+1}}.\label{fbra}
\ee In \eqref{sbra} and \eqref{fbra}, the subscripts $s$ and $f$ correspond to scalar branch and fermionic branch, respectively. 
 Since the fall-off index $\Delta$ of the fermionic theory differs from the power-law  exponent $\Delta+1$, we revisit the relation between the Carrollian amplitude and the momentum spacetime amplitude with fermionic incoming/outgoing states. This yields an extra factor of $\sqrt{\omega}$, as shown in \eqref{caf}. We discuss several properties of Carrollian correlators with fermionic operators and explicitly calculate some four-point Carrollian amplitudes in Yukawa theory and massless QED.
 By extrapolating the remaining bulk field to the boundary, we also investigate the resulting boundary-to-boundary correlators, uncovering several interesting properties summarized in table \ref{table3}. A double vertical line separates the three regimes defined by the value of the fall-off index $\Delta$. The symbol $\times$ indicates that the corresponding branch does not exist, $\checkmark$ indicates its existence, and $\infty$ signifies a divergent result that requires regularization.
 \begin{table}
\centering 
\renewcommand{\arraystretch}{1.5}
\begin{tabular}{||c|c|c||c|c|c||c|c|c||}\hline
$0<\Delta<1$&electric&magnetic &$\Delta=1$&electric &magnetic &$\Delta>1$&electric &magnetic\\\hline
scalar&$\times $&$\checkmark$&scalar&$\checkmark$&$\checkmark$&scalar&$\infty$&$\checkmark$\\\hline
fermion&$\times $&$\times$&fermion&$\checkmark$&$\times$&fermion&$\infty$&$\times$\\\hline
\end{tabular}
\caption{\centering{Boundary-to-boundary correlators in each branches}}
\label{table3}
\end{table}
 There are several issues that deserve further study. 
 \begin{itemize}
     \item \textbf{Theories without Poincar\'e symmetry.} The results for bulk-to-boundary correlators derived above rely on Poincar\'e symmetry. In a generic asymptotically flat spacetime, however, global Poincar\'e symmetry is absent in the bulk. Therefore, the correlators \eqref{sbra} and \eqref{fbra} may no longer hold, and the classification in table \ref{table3} could be altered due to bulk gravitational dynamics. 
      \item \textbf{Relation to K\"{a}ll\'{e}n-Lehmann representation}. The K\"{a}ll\'{e}n-Lehmann spectral representation \cite{Kallen:1952zz,Lehmann:1954xi} gives a general expression for the exact two-point function of interacting field theory as a sum of free propagators. For interacting scalar field theory
      \bea 
      G(x;x')=\int_0^\infty ds \rho(s)G(x;x';s) \label{klre}
      \eea where $G(x;x';s)$ is the scalar free propagator with $s=m^2$
      \bea 
      G(x;x';s)=\int \frac{d^4p}{(2\pi)^4}\frac{i}{-p^2-s+i\epsilon}e^{ip\cdot (x-x')}.
      \eea The parameter $s$ is positive and $\rho(s)$ is called the spectral function that is non-negative $\rho(s)\ge 0$ and obeys the sum rule 
      \be 
    1= \int_0^\infty \rho(s)ds.\label{sumrule}
      \ee The positivity of the spectral function follows from the unitarity of the theory. In a massive field theory, the momentum space propagator contains a pole at $p^2=-m^2$, representing the single particle contribution. The spectral function then takes the form
\be
\rho(s)=Z \delta(s-m^2)+\sigma(s),\label{rhos}
\ee
where $Z \le 1$ is a positive renormalization factor and $\sigma(s) \ge 0$ encodes the contributions from multi-particle states. Typically, a mass gap exists between the one particle state and the multi-particle continuum. Notice that for one particle sates $| k\rangle$,specified by a momentum $\bm k$,the energy is $\omega=\sqrt{\bm k^2+m^2}$, while for the multi-pariticle continumm $|\bm k,n \rangle$, specified by a momentum $\bm k $ and other parameter $n$, the energy of the state is $\omega=\sqrt{\bm k^2+M^2}$, where $M \gtrsim 2m$ is one of the parameters in the set $n$ \cite{2007qft..book.....S}. For massless theories, however, this is not guaranteed, as multi-particle states can also become massless, potentially leading to a non-unique vacuum. It thus appears necessary to distinguish between gapless systems and those with a mass gap. Note that the correlator $G(x;x')$ depends only on the Lorentz invariant interval $h=-(x-x')^2$ and can be expressed in the form given in \cite{Johnston:2009fr}
      \bea 
     G(x;x';s)= \frac{\sqrt{s}}{4\pi^2\sqrt{-h}}K_1(\sqrt{-s h}),\quad h<0
      \eea for spacelike interval.\footnote{For timelike distance, the correlator is proportional to the Hankel function of the second kind
      \be 
      G(x;x';s)=-\frac{i \sqrt{s} H_1^{(2)}\left(\sqrt{h s}\right)}{8 \pi  \sqrt{h }},\quad h>0.
      \ee } The function $K_1$ is the modified Bessel function of the second kind. 
      It is easy to check that as $m\to 0$, the above function becomes the free massless propagator 
      \be 
      \lim{}_{s\to 0}G(x;x';s)=\frac{1}{4\pi^2}\frac{1}{-h+i\epsilon}.
      \ee 
      Substituting into the K\"{a}ll\'{e}n-Lehmann representation \eqref{klre}, we find
      \bea 
      G(x;x')&=&\int_0^\infty ds \rho(s) \frac{\sqrt{s}}{4\pi^2\sqrt{-h}}K_1(\sqrt{s}\sqrt{-h})=\int_0^\infty ds'  \frac{\sqrt{s'}}{4\pi^2}\frac{\rho(\frac{s'}{-h})}{(-h)^2} K_1(\sqrt{s'}).\nn\\
      \eea Now we can extract the bulk-to-boundary correlator from K\"{a}ll\'{e}n-Lehmann representation by moving $x^\mu$ to $\mathcal I^+$. For gapless system, at the leading order, this is 
      \bea 
      G(x;x')=\frac{1}{4\pi^2}\int_0^\infty ds' \sqrt{s'} \rho\left(\frac{s'}{2r\widehat u}\right) \frac{1}{4r^2\widehat u^2 }K_1(\sqrt{s'})+\cdots.
      \eea On the other hand, the bulk-to-boundary correlator is fixed by Ward identity \eqref{bulktoboundary} and the fall-off behavior of $G(x;x')$ should be \footnote{ As has been discussed, a more general fall-off  condition \eqref{morefall} is also possible. In this case, the asymptotic expansion of the spectral function in the zero mass limit is 
      \be 
      \rho(s)=A s^{\Delta-2}\ln^\alpha s+\cdots,\quad s\to 0
      \ee and then normalization constant is fixed to 
      \be 
      C_s=\frac{A}{4\pi^2 2^\Delta}\int_0^\infty ds' s'^{\Delta-\frac{3}{2}}\ln^\alpha\frac{s'}{2}K_1(\sqrt{s'}).
      \ee }
      \be 
      G(x;x')=\frac{D(u,\Omega;x')}{r^\Delta}+\cdots.
      \ee This fixes the asymptotic behaviour of the spectral function  in the zero mass limit 
      \be 
      \rho(s)=A s^{\Delta-2}+\cdots,\quad s\to 0.\label{intrho}
      \ee The coefficient $A>0$ is related to the normalization constant $C_s$ \cite{2007tisp.book.....G}
      \bea 
      C_s=\frac{A}{4\pi^22^\Delta}\int_0^\infty ds' s'^{\Delta-\frac{3}{2}}K_1(\sqrt{s'})=\frac{A}{16\pi^2}2^\Delta\Gamma(\Delta)\Gamma(\Delta-1).\label{besselint}
      \eea Note that the asymptotic expansion \eqref{intrho} cannot satisfy the sum rule \eqref{sumrule} for $\Delta<1$ since the integral 
    $
      \int ds s^{\Delta-2}
      $ is divergent near $s=0$.
      The critical value $\Delta=1$ should be discussed separatedly and it can be saturated by free scalar with $\rho(s)=\delta(s)$. When there is a mass gap, then the spectral function may be written as 
      \be 
      \rho(s)=Z \delta(s)+\sigma(s)
      \ee where $\sigma(s)$ is non-vanishing only for $s>m_0^2$ with $m_0>0$ the mass gap. Then the dominant term near $\mathcal I^+$ is from the Dirac delta function and thus the fall-off index is $\Delta=1$ and 
      \be 
      C_s=\frac{Z}{8\pi^2}.\label{CsZ}
      \ee 
      
      The K\"{a}ll\'{e}n-Lehmann representation of interacting Dirac spinor propagator in a parity invariant theory can be written as \cite{1984PhT....37i..80I,2007qft..book.....S}
      \bea 
      G(x;x')&=&\int_0^\infty ds \int\frac{d^4p}{(2\pi)^4} \frac{i\left(\slashed p\rho_1(s)-\sqrt{s}\rho_2(s)\right)}{-p^2-s+i\epsilon}e^{ip\cdot (x-x')}\nn\\&=&-\int_0^\infty ds \left(i\rho_1(s)\slashed\partial +\sqrt{s}\rho_2(s)\right)G(x;x';s)
      \eea where the spectral functions $\rho_1(s)$ and $\rho_2(s) $ are both real  with $\rho_1(s)\ge 0$ and also larger than $\rho_2(s)$. The spectral functions should be expanded asymptotically as follows 
      \be 
      \rho_1(s)=A_1 s^{\Delta-2}+\cdots,\quad \rho_2(s)=A_2 s^{\Delta-\frac{5}{2}}+\cdots
      \ee to match with the fall-off behaviour \eqref{fallofffer} and the normalization constants of the bulk-to-boundary correlator are
      \bea 
      C_f=-\frac{iA_1}{16\pi^2}2^\Delta\Gamma(\Delta+1)\Gamma(\Delta-1),\quad C_s=-\frac{A_2}{16\pi^2}2^\Delta \Gamma(\Delta)\Gamma(\Delta-1).
      \eea  By examining the derivation of K\"{a}ll\'{e}n-Lehmann representation, we argue that in any unitary Poincar\'e invariant theory with a unique vacuum, the fall-off index $\Delta$ of the scalar or fermion field always saturates the bound $\Delta\ge 1$. In particular, for CFTs, the conformal dimension of a scalar primary field is equal to the fall-off index and unitarity bound is indeed $\Delta\ge 1$. For fermionic operator $\Psi$ in a CFT, the two point function is fixed to \be 
      \langle\text{T} \Psi(x)\bar\Psi(x')\rangle=\frac{\slashed x-\slashed x'}{(x-x')^{2\Delta_{\psi}+1}}
      \ee where the bound $\Delta_\psi\ge \frac{3}{2}$ follows from unitarity. The fall-off index is related to the  conformal weight $\Delta_\psi$  via 
      \be 
      \Delta=\Delta_\psi-\frac{1}{2}\quad\Rightarrow\quad \Delta\ge 1.
      \ee Thus, if this bound holds, the regime $\Delta < 1$ in table \ref{table3} should be excluded. However, the bound may be violated for non-gauge invariant operators as they are not strictly local observables.  We will discuss this point in the following.
      
     \item \textbf{Tensor fields.} It would be interesting to extend this analysis to bulk-to-boundary correlators involving tensor fields and check whether new branches emerge in such cases. A key challenge here is the gauge dependence of the gluon correlator. In QCD, non-perturbative gluon and ghost propagators are crucial for understanding quark confinement. The infrared exponent of the spectral function \cite{vonSmekal:1997ohs,Zwanziger:2001kw,Lerche:2002ep} governs the asymptotic behavior of these propagators, and one can explicitly verify that the aforementioned bound is violated. Furthermore, lattice simulations in Landau gauge indicate that the spectral function itself violates positivity \cite{Mandula:1987rh,Cucchieri:2004mf,Bowman:2007du}. Another intriguing direction involves bulk-to-boundary correlators for conserved currents and stress tensors. The K\"{a}ll\'{e}n-Lehmann representation for such conserved tensor fields are derived in general dimensions in \cite{Karateev:2020axc}. Their results suggest that we recover exactly the same number of independent structures for the bulk-to-boundary propagators across general dimensions\footnote{Working in progress.}. 
       \item \textbf{Bulk-to-timelike/spatial infinity correlators.}
     In our work, we have only considered bulk-to-boundary correlators where the boundary field is located at $\mathcal I^{\pm}$. For massive theories, the propagator to timelike infinity corresponds precisely to the external lines in perturbative QFTs \cite{Liu:2025oom}. One could also consider cases where the boundary field is placed at timelike or spatial infinity, and we expect such correlators to be similarly constrained by Poincar\'e symmetry. Moreover, the fall-off index of the bulk field near timelike/spatial infinity should be related to its counterpart near null infinity. The K\"{a}ll\'{e}n-Lehmann representation of scalar operator  \eqref{klre} can be written as 
     \bea 
     G(x;x')=\left\{\begin{array}{cc}-\frac{i}{8\pi h^2}\int_0^\infty ds' \sqrt{s'} \rho(\frac{s'}{h})H_1^{(2)}(\sqrt{s'})&h>0\\
     \frac{1}{4\pi^2 h^2}\int_0^\infty ds'\sqrt{s'}\rho(-\frac{s'}{h})K_1(\sqrt{s'})&h<0.\end{array}\right.
     \eea When the fall-off index at null infinity is $\Delta>1$, then the infrared behavior of the spectral function is \eqref{intrho}. As a consequence, the K\"{a}ll\'{e}n-Lehmann representation in the large $h>0$ expansion is 
   \begin{align} 
     G(x;x')\sim\left\{\begin{array}{cc}\frac{4^{\Delta} A(\cot (\pi  \Delta)-i) \Gamma (\Delta)}{16\pi  \Gamma (2-\Delta)}\frac{1}{h^\Delta}& h\to +\infty\\ \frac{4^{\Delta}A \Gamma (\Delta-1) \Gamma (\Delta)}{16\pi ^2}\frac{1}{(-h)^\Delta}&h\to-\infty.\end{array}\right.
     \end{align}

    \item \textbf{Conformal and non-conformal field theories.}
   Note that the analysis in this work is non-perturbative and it is still valid for strongly coupled systems. A QFT usually flows to a fixed point and the underlying theory becomes a CFT. It follows that the fall-off index should be equal to the conformal dimension (for scalar operators). However, there could be anomalous dimensions rather than naive classical dimensions. That means the fall-off index is corrected by quantum corrections. It is expected that these corrections can be reproduced by sum over loop corrections in Carrollian amplitude. It would be better to clarify this point in the future.  Another  interesting problem is to explore non-conformal field theories and determine the fall-off index. In  \cite{Billo:2017glv,Billo:2019job}, two-point correlators of chiral/anti-chiral operators in non-conformal
$\mathcal N=2$ supersymmetric Yang-Mills theories have been computed using perturbation theory and also localization method. The renormalized correlators  match with our fall-off conditions with possible logarithms. 
   \end{itemize}
  \vspace{3pt}
{\bf Acknowledgments.} 
The work of J.L. is supported by NSFC Grant No. 12575074.

\bibliography{refs}

\end{document}